\documentclass[11pt]{article}
\usepackage[colorlinks=true, citecolor=blue, urlcolor=blue, linkcolor=blue, breaklinks=true, pdfpagelabels=false]{hyperref}
\usepackage{cite}

\usepackage{hyperref}

\usepackage{amsmath,amsfonts,amssymb}
\usepackage{graphicx}
\usepackage{epsfig,epsf}
\usepackage{bm}
\usepackage{dsfont}
\usepackage{xcolor}

\def\a {\alpha}
\def\b {\beta}

\def\l {\lambda}

\def\bar {\overline}

\def\p {\partial}

\def\be {\begin{equation}}
\def\ee {\end{equation}}
\def\bea {\begin{eqnarray}}
\def\eea {\end{eqnarray}}
\def\n {\nonumber}

\def\barr{\begin{array}}
\def\earr{\end{array}}

\def\opcit(#1){ {\em op. cit.}, #1}

\def\issue(#1,#2,#3){#1, #2 (#3)} 

\def\equationautorefname~#1\null{Eq.\,(#1)\null}
\def\pageautorefname\nobreakspace{p.}

\makeatletter\renewcommand{\p@subsection}{\thesection.}\makeatother%
 
\begin{document}

\renewcommand*{\thefootnote}{\fnsymbol{footnote}}


\begin{center}
{\Large\bf{Oblique parameters of BSM models with three CP-even neutral scalars}}


\vspace{5mm}

{\bf Swagata Ghosh}$^{a,b,c}$\footnote{swgtghsh54@gmail.com}

\vspace{3mm}
{\em{${}^a$\ \ \ Department of Physics and Astrophysics, University of 
Delhi, Delhi, India.
}}

{\em{${}^b$\ \ \ SGTB Khalsa College, University of Delhi, Delhi, India.}}

{\em{${}^c$\ \ \ Department of Physics, Indian Institute of Technology Kharagpur, Kharagpur, India.
}}

\end{center}

\begin{abstract}
\noindent
We show the oblique parameters $\mathbb{S}$, $\mathbb{T}$, $\mathbb{U}$, $\mathbb{V}$, $\mathbb{W}$, and $\mathbb{X}$ as regards the corresponding mixing matrices in the frame of three BSM models with three CP-even neutral scalars. 
We look at three models with non-standard (i) two real singlet scalars, (ii) one complex doublet and one real singlet scalar, and (iii) two complex doublets. 
We handle the expressions similarly in that one can utilize these when all the neutral CP-even scalars own VEV or one of them doesn't own any VEV.
The top advantage of offering the oblique parameters in this way is, that the exclusive understanding of the mixing matrices in the particular scalar sector is sufficient to extract the phrasing of the oblique parameter of that individual BSM model. 
This paper also presents some plots, for reference, showing the allowed region of the masses of the three CP-even neutral scalars and the three mixing angles of the three above-mentioned models.
\end{abstract}




\setcounter{footnote}{0}
\renewcommand*{\thefootnote}{\arabic{footnote}}

\section{Introduction}
\label{intro}

\numberwithin{equation}{section}

To parametrise the upshots of new physics on the Electroweak precision observables, the oblique parameters play an important part \cite{Peskin:1991sw}.
This requires to satisfy some criteria, such that, $(i)$ The electroweak gauge group has to be the standard $SU(2)_L\times U(1)_Y$, with no non-standard gauge bosons, $(ii)$ New Physics particles combine to the light fermions restrainedly, although they combine to the gauge bosons in the main, $(iii)$ The new physics energy scales are at $q^2\approx 0$, $q^2=m_Z^2$, and $q^2=m_W^2$.
The relevant six quantities, viz, $\mathbb{S}$, $\mathbb{T}$, $\mathbb{U}$, $\mathbb{V}$, $\mathbb{W}$, $\mathbb{X}$ are established and expressed in \cite{Peskin:1991sw, 0802.4353ref1, 0802.4353ref2, 0802.4353ref3}.
Almost always, the SM participation in an oblique parameter $(\mathbb{O}_{SM})$ is taken off the NP contribution $(\mathbb{O}_{NP})$ to describe the oblique parameter $(\mathbb{O})$, $i.e.$
\be
 \mathbb{O}=\mathbb{O}_{NP}-\mathbb{O}_{SM}.
\ee
The $\mathbb{S}$ parameter captures the new physics contributions to the difference in vacuum polarization of the electroweak neutral gauge bosons ($W^3$ and $B$). 
In the SM, the $\mathbb{S}$ parameter is defined and computed purely from loops involving SM particles. 
Here, it possesses a small value ($0.00\pm0.07$), and often the reference point is set at $0$. 
As a result, any deviation from this value indicates the possibility of new physics. 
The $\mathbb{T}$ parameter describes electroweak radiative corrections to gauge-boson propagators, independent of vertex or box corrections. 
In the SM, the $\rho$ parameter at tree level remains identical to one, but for BSM models, its value differ from unity.
As a result, one can restrict any BSM model in such manner.
The oblique parameter $\mathbb{T}$, is connected to the $\rho$ parameter as,
\be
 \Delta{\mathfrak{\rho}}=\a \mathbb{T},
\ee
where, $\Delta\rho$ is the new physics part of the quantity, and $\a=e^2/(4\pi)$ is the fine structure constant.
The $\mathbb{U}$ parameter measures differences in the momentum dependence of the self-energies of the charged and neutral gauge-bosons. 
In the SM, $\mathbb{U}$ parameter at tree level remains $0$, and deviation of this value indicates the presence of BSM model. 
The $\mathbb{V}$ and $\mathbb{W}$ parameters estimate the momentum-dependent curvature of the $Z$-boson and $W$-boson propagator, respectively, beyond what is assessed by the oblique parameters $\mathbb{S,\,T},$ and $\mathbb{U}$. 
The $\mathbb{X}$ parameter amounts the nonlinearity of the mixing between the photon and the $Z$-boson with respect to the energy. 

Amid several non-standard models, inclusion of BSM particles under some symmetries are of distinct scrutiny. 
Adjunct of the scalar sector by one or more, real or complex, charged or neutral multiplets are well deduced in published works, for example \cite{Barger:2007im, Barger:2008jx,
Ahriche:2013vqa, Branco:2011iw, Grossman:1994jb}. 
This work does not focus on any extra charged scalar as no Higgs-like physics is reproduced from it. 
Any BSM model containing more than one CP-even neutral scalar, must possess the Higgs-like scalar consistent with the present collider data.
Though the presence of two CP-even neutral scalars in the scalar sector of any BSM model is sufficient for the minimal extension, the presence of three CP-even neutral scalars provides minimal non-trivial scalar structure which may address a richer phenomenological framework. 
This paper examines three sets of addition of the scalar sector of the SM, viz,
$(i)$ SM extended with two real singlet scalars (Rx2SM) \cite{Robens:2019kga}, $(ii)$ SM extended with scalars of which, one is complex doublet and another one is real singlet (N2HDM), and $(iii)$ SM extended with two complex doublet scalars (3HDM).
All of these BSM models possess three neutral scalars, which are CP-even in nature. 
A general expression of the oblique parameters for BSM models accompanied by multiple $SU(2)_L$ singlet and doublet are given in \cite{Grimus:2007if, Grimus:2008nb}. 
We use these results in our paper to provide the expressions of the oblique parameters as to the rotation matrices and few preconceived functions.

The six oblique parameters $\mathbb{S,\,T,\,U,\,V,\,W,\,{\rm and}\,X}$ relate differently to the three above mentioned BSM models. 
In the Rx2SM, the $\mathbb{S}$ parameter arises only from mixing of singlets with the SM Higgs and is typically small unless mixing is large or masses are widely split. 
No direct gauge couplings from singlets make $\mathbb{S}$ naturally suppressed. 
In the N2HDM, the $\mathbb{S}$ parameter can be sizable due to additional contributions from the doublet sector, especially if CP-even or CP-odd scalars are split in mass. 
Singlet effects are indirect, but the 2HDM structure gives this model more flexibility to impact $\mathbb{S}$. 
In the 3HDM, the $\mathbb{S}$ parameter can be large due to multiple doublets contributing directly to gauge boson loops, especially with non-degenerate scalar masses. 
It is one of the most constrained parameters in this model from precision electroweak data. 
In the Rx2SM, the $\mathbb{T}$ parameter reflects custodial symmetry breaking due to unequal mixing and mass splitting among scalars. 
Even though singlets are neutral, their mixing with the Higgs can introduce effective isospin-breaking. 
In the N2HDM, the $\mathbb{T}$ parameter is very sensitive to the scalar spectrum, especially mass differences between charged Higgs, CP-even, and CP-odd scalars. 
Custodial symmetry is more easily broken in this model, making $\mathbb{T}$ a key constraint. 
In the 3HDM, the $\mathbb{T}$ parameter can be significantly violated if care isn't taken to preserve custodial symmetry, due to many charged and neutral scalars. 
Mass splittings and vev alignments play a crucial role in determining its size. 
In the Rx2SM, the $\mathbb{U}$ parameter is usually negligible due to the absence of charged scalars or $SU(2)$ multiplet structure. 
It is typically consistent with zero unless derivative interactions are introduced. 
In the N2HDM, the $\mathbb{U}$ parameter can receive small contributions from the 2HDM sector but is still generally subdominant compared to $\mathbb{S}$ and $\mathbb{T}$ parameters. 
Its impact is marginal unless scalar couplings have strong momentum dependence. 
In the 3HDM, the $\mathbb{U}$ parameter may become non-negligible if scalar mass splittings are large and charged-neutral asymmetry exists. 
Still, it's usually suppressed and less constraining than $\mathbb{S}$ as well as $\mathbb{T}$. 
In the Rx2SM, the $\mathbb{V}$ parameter is typically small and arises only from energy-dependent $Z$-boson self-energy corrections via scalar loops. 
It is insignificant unless there's strong scalar mixing and high mass hierarchy. 
In the N2HDM, the $\mathbb{V}$ parameter can be moderately affected by the 2HDM scalars if mass gaps are large. 
It's more relevant near the $Z$-pole or higher-energy processes. 
In the 3HDM, the $\mathbb{V}$ parameter can be enhanced due to the dense scalar spectrum and multiple interactions affecting the $Z$-propagator. 
Still, precision fits typically keep it under control unless high-energy tails are probed. 
In the Rx2SM, the $\mathbb{W}$ parameter is minimal due to lack of charged scalars and absence of direct coupling to $W$ bosons. 
It remains almost zero in most benchmark scenarios. 
In the N2HDM, the $\mathbb{W}$ parameter can receive contributions from charged Higgs loops in 2HDM. 
However, unless mass splittings are large, it's still smaller than the $\mathbb{T}$ parameter. 
In the 3HDM, the $\mathbb{W}$ parameter may be enhanced due to multiple charged scalars contributing to $W$-boson propagators. 
It's especially relevant when scalar mass spectrum is broad and hierarchical. 
In the Rx2SM, the $\mathbb{X}$ parameter is essentially zero since singlets are neutral and cannot induce $Z\gamma$ mixing at one loop. 
Only indirect effects through Higgs mixing exist, which are negligible. 
In the N2HDM, the $\mathbb{X}$ parameter is generally small but can arise from loops involving the charged Higgs in 2HDM. 
It's usually a minor correction in precision fits. 
In the 3HDM, the $\mathbb{X}$ parameter can be larger if there are multiple charged scalars with asymmetric couplings to $Z$ and $\gamma$. 
Still, it remains subleading unless there's significant charge and mass asymmetry. 

For BSM models containing additional non-standard multiplets, larger than $SU(2)_L$ doublets, $\mathbb{S}$ and $\mathbb{U}$  are prescribed in  \cite{Albergaria:2021dmq}.
For aligned $2HDM$ and $3HDM$, oblique parameters are specified in \cite{Jurciukonis:2021wny}.
Some formulations of some oblique parameters in these models are already in literature \cite{Branco:2011iw,Farzinnia:2013pga,Chao:2016cea,Darvishi:2016gvm,Belanger:2014bga,ONeil:2009fty,Haber:2010bw,Swiezewska:2012ej,Davidson:2010xv,Botella:2014ska,Aoki:2009ha,Boucenna:2011hy,Grzadkowski:2009bt,Kanemura:2015mxa,Ferreira:2011xc,Swiezewska:2012ej,Melfo:2011ie,Moretti:2015cwa,deAdelhartToorop:2010jxh,Boucenna:2011tj,aali:2020tgr,Arhrib:2018qmw}, yet all the six oblique parameters $(\mathbb{S, T, U, V, W, X})$ concerning the elements of the rotation matrices, for the two real singlet scalar extended SM, one real singlet scalar extended 2HDM, and three Higgs doublet model, are not expressed in previous works.
Our chief goal for this study is imparting the complete list of all the six oblique parameters for the above-mentioned BSM models.

This work is arranged thusly.
Section \ref{model} contains the brief description of the models.
Section \ref{Results} enlists the oblique parameters for these three models.
We conclude in Section \ref{Conclusions}.
Mixing matrices, and the calculation in detail may be found in the appendices \ref{A} and \ref{B}.

\section{The models}
\label{model}

\numberwithin{equation}{subsection}

This section gives a brief description of the scalar sector of some BSM models, extended by singlet(s) or/and doublet(s) in addition to the SM doublet. 
The subsection \ref{GeneralModel} briefly provides the general method to write the required matrices to express the oblique parameters in terms of some functions and the components of the rotation matrices between the weak eigenstates and the mass eigenstates. 
This subsection consider the extension of the scalar sector of the SM by $(n_D-1)$ no. of complex doublets and $n_R$ no. of real singlet scalars. 
The extension of the scalar sector of the SM by two real singlet scalars is given in \ref{Rx2SM model}. 
No addition of the complex doublet scalar is considered in this model. 
The next subsection \ref{N2HDM model} contemplates the inclusion of one complex doublet and one real singlet scalar to the SM scalar sector. 
The incorporation of two additional complex doublet is evaluated in the succeeding subsection \ref{3HDM model}. 
No extra singlet is introduced in the scalar sector of this model. 
\subsection{The scalar extended Standard Model}
\label{GeneralModel}
First, following \cite{Grimus:2007if,Grimus:2008nb}, let us consider an $SU(2)_L\times U(1)$ BSM model in a general way, before treating any specific individual model. 
The scalar sector of this model may contain $n_D$ no. of $SU(2)_L$ doublets ($Y=1/2$) and $n_R$ no. of real singlet scalars ($Y=0$) of $SU(2)_L$, which are given by, respectively, 
\begin{equation}
 \Phi_j = 
 \begin{pmatrix} 
  \phi_j^+ \cr \phi_j^0 
 \end{pmatrix}\,\, (j=1,2,..,n_D)\,,
 \qquad
 \chi_k \,\, (k=1,2,..,n_R)\,.
\end{equation}
After acquiring VEVs, the neutral fields become :
\begin{equation}
 \phi_j^0 = \frac{v_j+\phi_j^{0^{\prime}}}{\sqrt{2}}\,, \qquad
 \chi_k = u_k + \chi_k^{\prime}\,,
 \label{eq:fields}
\end{equation}
such that, the electroweak VEV is expressed as $v = \big(\sum_{j=1}^{n_D} |v_j|\big)^{1/2} \simeq 246$ GeV. 
Therefore, the model possesses $m = (2 n_D + n_R)$ no. of real neutral scalar fields and $n_D$ no. of singly charged complex scalar fields. 
The charged and neutral mass-eigenstates are then respectively represented by $S_a^+$ and $S_b^0$, with ($a=2,..,n_D$) and ($b=2,..,m$). 
$S_1^{\pm}$ and $S_1^0$ are assigned to the three Goldstone bosons $G^{\pm},\,G^0$ respectively. 
The components of the doublets and the singlets can be expressed as, 
\begin{equation}
 \phi_j^+ = \sum_{a=1}^{n_D} {\cal{U}}_{ja} S_a^+\,,
 \qquad
 \phi_j^0 = \frac{v_j + \sum_{b=1}^m {\cal{V}}_{jb} S_b^0}{\sqrt{2}}\,,
 \qquad
 \chi_k = u_k + \sum_{b=1}^m {\cal{R}}_{kb} S_b^0\,.
 \label{eq:components}
\end{equation}
The dimensions of the matrices ${\cal{U}},\, {\cal{V}},\, {\cal{R}}$ are given by, $n_D \times n_D$, $n_D \times m$, $n_R \times m$, respectively. 
In the SM, $n_D=1$, $n_R=0$, and $m=2$ to give ${\cal{U}} = {\mathds{1}}$, ${\cal{V}} = (i\,\,\,1)$. 
The matrix $\cal{R}$ does not exist in the SM. 

In the next subsections, 
we shortly recall the most general scalar sectors of the models used in our paper, where, all neutral scalars have VEVs, and as a result, they mix with each other. 
We present the models accordingly.
For the cases, where all the neutral scalars do not own VEV, the mixing matrices are reduced.

\subsection{The Two Real Singlet Scalar extended Standard Model (Rx2SM)}
\label{Rx2SM model}

Concerning the component fields 
of the scalar sector of the Rx2SM,
 $\Phi = 
  \Big(\phi^{+} \,\, \phi^{0}\Big)^{T}
 $,
 $\chi_1^{}
 $,
 $\chi_2^{}
 $,
where $\chi_1^{}$ and $\chi_2^{}$ are the real singlets with hypercharge $Y = 0$ and $\Phi$ is the SM complex doublet with hypercharge $Y = 1$. 
\\
One can write the scalar potential as :
\bea
 V(\Phi_1,\chi_1,\chi_2) &=& 
 \mu_{\Phi}^2 \Phi^{\dag} \Phi \,+\, \mu_{\chi_1}^2 \chi_1^2 \,+\, \mu_{\chi_2}^2 \chi_2^2 \,+\, \l_{\Phi} (\Phi^{\dag} \Phi)^2 \,+\, \l_{\chi_1} \chi_1^4 \,+\, \l_{\chi_2} \chi_2^4 \nonumber\\
 &+& 
 \l_{\Phi\chi_1} \Phi^{\dag} \Phi \chi_1^2 \,+\, \l_{\Phi\chi_2} \Phi^{\dag} \Phi \chi_2^2 \,+\, \l_{\chi_1\chi_2} \chi_1^2 \chi_2^2
\label{eq:pot_Rx2SM}
\eea
Here, $\mu_{\Phi,\chi_1^{},\chi_2^{}}$ are of dimension $2$, and $\l_{\Phi}$, $\l_{\chi_1}$, $\l_{\chi_2}$, $\l_{\Phi\chi_1}$, $\l_{\Phi\chi_2}$, $\l_{\chi_1\chi_2}$ are dimensionless coefficients. 
All the parameters are real. 
This scalar potential is symmetric under the $Z_2 \otimes Z_2^{\prime}$, where under $Z_2$ symmetry $\Phi \rightarrow \Phi$, $\chi_1^{} \rightarrow -\chi_1^{}$, $\chi_2^{} \rightarrow \chi_2^{}$ and under $Z_2^{\prime}$ symmetry $\Phi \rightarrow \Phi$, $\chi_1^{} \rightarrow \chi_1^{}$, $\chi_2^{} \rightarrow -\chi_2^{}$. 
\\
After spontaneous symmetry breaking (SSB), the vacuum expectation value (VEV) for the singlets are $<\chi_1^{}>=u_1$ and $<\chi_2^{}>=u_2$, and that for the neutral field of the doublet is $<\phi^0>=v/\sqrt{2}$, where, $v$ is the electroweak VEV. 
These singlet VEVs help to realize different phases of this model. 
If either of the singlet VEVs is $0$, the corresponding singlet is the dark matter candidate, provided there is some symmetry in the potential, and the model is in dark phase. 
For $u_{1,\,2} \ne\,0$, all the three unphysical fields $\rho_0,\, \rho_s,\, \rho_x$ mix to provide the physical fields $H_{1,2,3}$, and the model is in broken phase. 
\\
\\
The neutral fields about their VEVs gives,
\bea
 \phi^0 &=& 
  \frac{1}{\sqrt{2}}(v+\phi^{0^{\prime}})\,
  \qquad {\rm with}\,\,
  \phi^{0^{\prime}}= \rho_0+i\,G^0\,, \nonumber\\
 \chi_1^{} &=& 
  u_1 + \chi_1^{\prime}\,
  \qquad\qquad {\rm with}\,\, 
  \chi_1^{\prime}= \rho_s\,, \nonumber\\
 \chi_2^{} &=& 
  u_2 + \chi_2^{\prime}\,
  \qquad\qquad{\rm with}\,\, 
  \chi_2^{\prime} = \rho_x\,.
\label{eq:phichiRx2SM}
\eea  
\\
The minimization conditions of the scalar potential \ref{eq:pot_Rx2SM} eliminate the mass terms as :
\bea
 \mu_{\Phi}^2 &=& - \l_{\Phi} v^2 - \l_{\Phi\chi_1} u_1^2 - \l_{\Phi\chi_2} u_2^2 \nonumber\\ 
 \mu_{\chi_1}^2 &=& - 2 \l_{\chi_1} u_1^2 - \frac12 \l_{\Phi\chi_1} v^2 - \l_{\chi_1\chi_2} u_2^2 \nonumber\\ 
 \mu_{\chi_2}^2 &=& - 2 \l_{\chi_2} u_2^2 - \frac12 \l_{\Phi\chi_2} v^2 - \l_{\chi_1\chi_2} u_1^2 
\label{eq:eliminate_Rx2SM}
\eea
\\
This paper only pays interest in the neutral CP-even scalars $\rho_0,\,\rho_s,\,\rho_x$ and hence only the corresponding mass-matrix is given here, as : 
\bea
 {\cal{M}}_{CP-even}^2 = 
 \begin{pmatrix}
  2 \l_{\Phi} v^2 & 2 \l_{\Phi\chi_1} u_1 v & 2 \l_{\Phi\chi_2} u_2 v \cr
  2 \l_{\Phi\chi_1} u_1 v & 8 \l_{\chi_1} u_1^2 & 4 \l_{\chi_1\chi_2} u_1 u_2 \cr
  2 \l_{\Phi\chi_2} u_2 v & 4 \l_{\chi_1\chi_2} u_1 u_2 & 8 \l_{\chi_2} u_2^2
 \end{pmatrix}
\label{eq:MassMatrix_Rx2SM}
\eea
\\
This model contains six scalars, out of which three are Goldstone bosons, and the remaining three charge-neutral scalars are CP-even in nature. 
One can secure the mass eigenstates from the unphysical fields, through the rotation matrix, as,
\be
 \begin{pmatrix}
  H_1 \cr H_2 \cr H_3 
 \end{pmatrix} 
 = O_{\a} 
 \begin{pmatrix}
  \rho_0 \cr \rho_s \cr \rho_x
 \end{pmatrix}\,, \qquad {\rm where}\,\, O_{\a}\,\, {\rm is\,\, a }\,\, 3\times3 \,\,{\rm orthogonal\,\, matrix}\,.
\ee
The hierarchy of the masses of the CP-even scalars is assumed here such that, 
$m_{H_1}\le m_{H_2} \le m_{H_3}$, 
where $m_{H_i}$ is the mass of $H_i$. 
One of these three Higgs bosons must be considered to be the SM Higgs boson. 
The scalar $H_2$ is treated here as the observed SM Higgs boson with mass at about $125$ GeV. 
The matrix $O_{\a}$ (see Appendix \ref{A}) diagonalises the mass matrix (Eqn. \ref{eq:MassMatrix_Rx2SM}) of the CP-even Higgs sector. 
For the broken phase, this rotation matrix is most general, and is given in the Appendix \ref{eq:Oalpha}. 
For the two dark phases, this matrix acquires two different forms. 
If $u_1 = 0$, the unphysical field $\rho_s$, considered as the dark matter candidate, does not mix with $\rho_0,\, \rho_x$ and has no couplings to SM particles. 
The corresponding rotation matrix is given in the Appendix \ref{eq:Oalpha1}. 
If $u_2 = 0$, the unphysical field $\rho_x$, considered as the dark matter candidate, does not mix with $\rho_0,\, \rho_s$ and has no couplings to SM particles. 
The corresponding rotation matrix is given in the Appendix \ref{eq:Oalpha2}. 
Further discussions on these phases are not posited here as these are out of scope of this paper. 
\\
\\
Now, to focus on the couplings between the Higgs and the fermions, one needs to consider the Yukawa Lagrangian. 
In this model, the Higgs couplings to the fermions with mass $m_f$ are not as simple as that in the SM, where the coupling is $m_f/v$. 
One can write the Yukawa Lagrangian as, 
\be
 {\cal{L}}_{Yukawa}^{Rx2SM} = 
 - \Big[ 
 \overline{Q}_L \widetilde{\Phi} {\cal{Y}}_u u_R 
 \,+\, \overline{Q}_L \Phi {\cal{Y}}_d d_R 
 \,+\, \overline{L}_L \Phi {\cal{Y}}_l l_R
 \,+\, {\rm h.c.} \Big]\,,
\label{eq:Yukawa_Rx2SM}
\ee
where, $Q_L,\,L_L$ represent the quark doublet and the lepton doublet, while $u_R,\,d_R,\,l_R$ represent the singlets of up-type quark, down-type quark and lepton, respectively. 
${\cal{Y}}_{u,d,l}$ are the matrices in the flavor space, and $\widetilde{\Phi} \equiv i\sigma_2\Phi^{{}^{\star}}$. 
Using the components of the unphysical fields, one can rewrite the Eqn. \ref{eq:Yukawa_Rx2SM} in terms of the physical scalars ($\mathtt{s} =$ $H_1,\,H_2,\,H_3$), the fermions ($f=$ $u$, $d$, $l$), and the corresponding couplings $g_{\mathtt{s}}^f = (m_f/v)\,\xi_{\mathtt{s}}^f$, as given below. 
\be
 {\cal{L}}_{Yukawa}^{Rx2SM} = 
 - \sum_{f=u,d,l} \frac{m_f}{v} \Big( \xi_{H_1}^f \bar{f} f H_1 \,+\, \xi_{H_2}^f \bar{f} f H_2 \,+\, \xi_{H_3}^f \bar{f} f H_3 \Big)\,, 
\label{eq:xi_N2HDM}
\ee
with 
\be
 \xi_{H_a}^f = O_{\a_{a1}}\,,
\label{eq:Hff_Rx2SM}
\ee 
where, $O_{\a_{a_1}}$ are the components of the first column of the rotation matrix $O_{\a}$ given in the Appendix \ref{A}, for $(a=1,\,2,\,3)$. 
Clearly the Yukawa couplings in this model scaled to the same in the SM are denoted by $\xi_{H_a}^f$. 
\\
\\
The kinetic part of the Lagrangian of Rx2SM can be expressed in terms of the fields $\Phi,\,\chi_1^{},\,\chi_2^{}$ as : 
\be
 {\cal{L}}_{kin}^{Rx2SM} = 
 (D^{\mu} \Phi)^{\dag} (D_{\mu} \Phi) \,+\, \frac12 (\p^{\mu} \chi_1^{}) (\p_{\mu} \chi_1^{}) \,+\, \frac12 (\p^{\mu} \chi_2^{}) (\p_{\mu} \chi_2^{})\,, 
\label{eq:Lkin_Rx2SM}
\ee
which gives the ratio of couplings of the CP-even neutral scalars $H_{1,2,3}$ with the gauge bosons pair $VV = WW,\,ZZ$ ($g_{H_{1,2,3}}^{VV}$) to the same coupling in the SM ($g_{H_{SM}}^{VV}$), as, 
\be
C_{H_{1,2,3}}^{VV} = O_{\a_{a1}}
\label{eq:HVV_Rx2SM}
\ee
equal to the $\xi_{H_{1,2,3}}^f$, given in the Eqn. \ref{eq:Hff_Rx2SM}. 
\\
In general, the three unphysical fields can be expressed in terms of the physical fields and the components $(O_\a)_{pq}$, with ($p,q = 1,2,3$) of the rotation matrix as, 
\bea
 \rho_0 &=& 
 O_{\alpha_{11}} H_1 + O_{\alpha_{21}} H_2 + O_{\alpha_{31}} H_3 \nonumber\\
 \rho_s &=& 
 O_{\alpha_{12}} H_1 + O_{\alpha_{22}} H_2 + O_{\alpha_{32}} H_3 \nonumber\\
 \rho_x &=& 
 O_{\alpha_{13}} H_1 + O_{\alpha_{23}} H_2 + O_{\alpha_{33}} H_3 
\label{eq:rhoRx2SM}
\eea
Probing Eqn.\ref{eq:rhoRx2SM} into Eqn.\ref{eq:phichiRx2SM}, one can write the neutral components of the doublet and the singlets as, 
\bea
 \phi^0 &=& \frac{1}{\sqrt{2}} (v + i\,G^0 + O_{\alpha_{11}} H_1 + O_{\alpha_{21}} H_2 + O_{\alpha_{31}} H_3)\,, \nonumber\\
 \chi_1^{} &=& u_1 + O_{\alpha_{12}} H_1 + O_{\alpha_{22}} H_2 + O_{\alpha_{32}} H_3\,, \nonumber\\ 
 \chi_2^{} &=& u_2 + O_{\alpha_{13}} H_1 + O_{\alpha_{23}} H_2 + O_{\alpha_{33}} H_3\,.
 \label{eq:neutralRx2SM}
\eea
\\
We assign the physical charged and neutral scalars respectively, as,
\begin{equation}
 S^{+}=G^{+}\,,
 S_1^0=G^0,\,\,S_2^0=H_1,\,\,S_3^0=H_2,\,\,S_4^0=H_3.
 \label{eq:fieldsRx2SM}
\end{equation}
\\
We have $n_D=1$ charged scalar field, $S^+$, which is Goldstone boson, and $m=4$ real neutral scalar fields, $S_{1,2,3,4}^0$, out of which $S_1^0$ is the Goldstone boson.
The charged scalar $S^+$ is affixed to the unphysical scalar $\phi^+$ through the matrix ${\cal{U}}$.
The matrices ${\cal{V}}$ and ${\cal{R}}$ associate with the neutral component of the SM doublet $\Phi$ and that of the BSM singlets $\chi_{1,2}^{}$ to the physical neutral scalars $S_{1,2,3,4}^0$ respectively.
Following the procedures described in the general section \ref{GeneralModel}, in terms of the matrices $\cal{U,\,V,\,R}$ and the newly defined scalars $S^+$ and $S_{1,2,3,4}^0$ as given in the Eqn. \ref{eq:fieldsRx2SM}, the components of the doublet $\Phi$ and the singlets $\chi_1^{},\,\chi_2^{}$ can be expressed as, 
\bea
 \phi^{+}= {\cal{U}} S^+,\quad
 \phi^{0^{\prime}}=\sum_{b=1}^4 {\cal{V}}_{1b}S_b^0,\quad
 \chi_k^{\prime}=\sum_{b=1}^4 {\cal{R}}_{kb}S_b^0\,,\quad (k=1,2), 
\label{eq:compmatRx2SM} 
\eea
where the matrices ${\cal{U}}$, ${\cal{V}}$ and ${\cal{R}}$ are of dimensions $1\times1$, 
$1\times4$ and $2\times4$ respectively. 
\\
Also, relating the Eqns. \ref{eq:phichiRx2SM} and \ref{eq:neutralRx2SM}, one can write the components $\phi^+,\,\phi^{0^{\prime}},\,\chi_1^{\prime},\,{\rm and}\,\chi_2^{\prime}$ in terms of the components of the mixing matrix $O_{\a}$ and the scalars defined in the Eqn. \ref{eq:fieldsRx2SM} as, 
\bea
 \phi^+ &=& S^+\,, \nonumber\\
 \phi^{0^{\prime}} &=& 
 i\,S_1^0 + O_{\alpha_{11}} S_2^0 + O_{\alpha_{21}} S_3^0 + O_{\alpha_{31}} S_4^0\,, \nonumber\\
 \chi_1^{\prime} &=& 
 O_{\alpha_{12}} S_2^0 + O_{\alpha_{22}} S_3^0 + O_{\alpha_{32}} S_4^0\,, \nonumber\\
 \chi_2^{\prime} &=&
 O_{\alpha_{13}} S_2^0 + O_{\alpha_{23}} S_3^0 + O_{\alpha_{33}} S_4^0\,.
\label{eq:phichiprimeRx2SM}
\eea
Therefore, comparing the Eqns. \ref{eq:phichiprimeRx2SM} with \ref{eq:compmatRx2SM} the matrices
are given by,
\bea
 {\cal{U}}=\mathds{1}\,, \quad
 {\cal{V}}=
  \begin{pmatrix}
  i & O_{\a_{11}} & O_{\a_{21}} & O_{\a_{31}}
 \end{pmatrix}\,,\quad
 {\cal{R}}=
 \begin{pmatrix}
  0 & O_{\a_{12}} & O_{\a_{22}} & O_{\a_{32}} \cr
  0 & O_{\a_{13}} & O_{\a_{23}} & O_{\a_{33}} \,
 \end{pmatrix}\,.
 \label{eq:UVR1}
\eea

\subsection{The Real Singlet Scalar extended Two Higgs Doublet Model (N2HDM)}
\label{N2HDM model}

The scalar sector of this model consists of two complex $SU(2)_L$ doublets with hypercharge $Y = 1$ and one real $SU(2)_L$
singlet with hypercharge $Y = 0$, concerning the component fields,
$ \Phi_j = 
 \Big(\phi_j^{+} \,\, \phi_j^{0}\Big)^T$,
 with j = 1,2\,, 
and
$ \chi$,
where $\chi$ is the singlet and $\Phi_j$ is the $j$-th doublet. 
\\
One can write the scalar potential as :
\bea
 V(\Phi_1,\Phi_2,\chi) &=& 
 \mu_{11}^2 \Phi_1^{\dag} \Phi_1 \,+\, \mu_{22}^2 \Phi_2^{\dag} \Phi_2 \,-\, \mu_{12}^2 (\Phi_1^{\dag} \Phi_2 + \Phi_2^{\dag} \Phi_1) \,+\, \frac12 \mu_{\chi}^2 \chi^2 \nonumber\\
 &+& 
 \frac{\l_1}{2} (\Phi_1^{\dag} \Phi_1)^2 \,+\, \frac{\l_2}{2} (\Phi_2^{\dag} \Phi_2)^2 \,+\, \l_3 \Phi_1^{\dag} \Phi_1 \Phi_2^{\dag} \Phi_2 \,+\, \l_4 \Phi_1^{\dag} \Phi_2 \Phi_2^{\dag} \Phi_1 \nonumber\\
 &+& 
 \frac{\l_5}{2} \Big[ (\Phi_1^{\dag} \Phi_2)^2 \,+\, (\Phi_2^{\dag} \Phi_1)^2 \Big] \,+\, \frac{\l_6}{8} \chi^4 \,+\, \frac12 \Big[ \l_7 \Phi_1^{\dag} \Phi_1 + \l_8 \Phi_2^{\dag} \Phi_2 \Big] \chi^2\,.
\label{eq:pot_N2HDM}
\eea
Here, $\mu_{11,22,12,\chi}$ are of dimension $2$, and $\l_{1-8}$ are dimensionless coefficients. 
The parameters $\mu_{11}$, $\mu_{22}$, $\mu_{\chi}$, $\l_1$, $\l_2$, $\l_3$, $\l_4$, $\l_6$, $\l_7$, $\l_8$ are real while $\mu_{12}$, $\l_5$ can be complex, but this paper assumes all these parameters to be real \cite{Arhrib:2018qmw}. 
This scalar potential is symmetric under the $Z_2 \otimes Z_2^{\prime}$, where under $Z_2$ symmetry $\Phi_1 \rightarrow \Phi_1$, $\Phi_2 \rightarrow -\Phi_2$, $\chi \rightarrow \chi$ and under $Z_2^{\prime}$ symmetry $\Phi_1 \rightarrow \Phi_1$, $\Phi_2 \rightarrow \Phi_2$, $\chi \rightarrow -\chi$. 
The $Z_2$ symmetry is softly broken by the presence of the term $\mu_{12}$. 
\\
After spontaneous symmetry breaking (SSB), the vacuum expectation value (VEV) for the singlet is $<\chi>=u$, and that 
for $j$-th neutral field is $<\phi_j^{0}>=v_j/\sqrt{2}$, such 
that, the total electroweak VEV, $v$, can be expressed as,
$
 v^2=\sum_{j=1}^2 v_j^2 = (246 \rm\, GeV)^2.
$
\\
After expanding the neutral fields about their VEVs, 
\bea
 \phi_j^0 &=& 
  \frac{1}{\sqrt{2}}(v_j +  \phi_j^{0^{\prime}})\,, 
  \qquad{\rm with}\,\, 
  \phi_j^{0^{\prime}} = \rho_j + i\,\eta_j\,, \nonumber\\
 \chi &=& u + \chi^{\prime} 
  \,, 
  \qquad\qquad\quad{\rm with}\,\, \chi^{\prime}= \rho_s\,. 
\label{eq:phichiN2HDM}
\eea 
The two doublet VEVs can be expressed in terms of the mixing angle $\b$ such that, $v_1=v \cos{\b}$ and $v_2=v\sin{\b}$, leading to the ratio of the doublet VEVs 
\be
\frac{v_2}{v_1} = \tan{\b}.
\label{eq:tanb_N2HDM}
\ee
\\
The VEVs help to realize different phases of this model. 
If either of the doublet VEVs $v_j$ is zero, the corresponding real part of the neutral unphysical field $\rho_j$ of the doublet $\Phi_j$ is the dark matter candidate, provided there is some symmetry in the potential, and the model is in dark phase. 
This dark matter candidate does not mix with other unphysical scalars and has no couplings to the SM particles. 
Similarly, if the singlet VEV $u$ is $0$, the unphysical field $\rho_s$ of the singlet $\chi$ is the dark matter candidate for the existance of some symmetry in the potential, and the model is again in dark phase. 
$\rho_s$ does not mix with $\rho_{1,2}$ in this case, and has no couplings to the SM particles. 
For $u,\,v_1,\,v_2\ne0$, all the three unphysical fields $\rho_1,\,\rho_2,\,\rho_s$ mix to provide the physical fields $H_{1,2,3}$ resulting the broken phase. 
This paper does not consider these phases further, as those are out of scope of this discussion. 
\\\\
The minimization conditions of the scalar potential \ref{eq:pot_N2HDM} eliminate the mass terms as :
\bea
 \mu_{11}^2 &=& \mu_{12}^2 \tan{\b} - \frac12 \l_1 v^2 \cos^2{\b} - \frac12 (\l_3+\l_4+\l_5) v^2 \sin^2{\b} - \frac12 \l_7 u^2 \nonumber\\ 
 \mu_{22}^2 &=& \mu_{12}^2 \tan^{-1}{\b} - \frac12 \l_1 v^2 \sin^2{\b} - \frac12 (\l_3+\l_4+\l_5) v^2 \cos^2{\b} - \frac12 \l_8 u^2 \nonumber\\ 
 \mu_{\chi}^2 &=& -\frac12 \l_7 v^2 \cos^2{\b} - \frac12 \l_8 v^2 \sin^2{\b} -\frac12 \l_6 u^2 
\label{eq:eliminate_N2HDM}
\eea
\\
This paper only pays interest in the neutral CP-even scalars $\rho_1,\,\rho_2,\,\rho_s$ and hence only the corresponding mass-matrix is given here, as : 
\bea
 {\cal{M}}_{CP-even}^2 = 
 \begin{pmatrix}
  \mu_{12}^2 \tan{\b} + \l_1 v^2 \cos^2{\b} & -\mu_{12}^2 + (\l_3+\l_4+\l_5) v^2 \sin{\b} \cos{\b} & \frac{\l_7}{2} u v \cos{\b} \cr
  -\mu_{12}^2 + (\l_3+\l_4+\l_5) v^2 \sin{\b} \cos{\b} & \mu_{12} \tan^{-1}{\b} + \l_2 v^2 \sin^2{\b} & \frac{\l_8}{2} u v \sin{\b} \cr
  \frac{\l_7}{2} u v \cos{\b} & \frac{\l_8}{2} u v \sin{\b} & \l_6 u^2
 \end{pmatrix}
\label{eq:MassMatrix_N2HDM}
\eea
\\
This model contains nine scalars, out of which three are Goldstone bosons $G^{0,\pm}$. 
N2HDM also contains three CP-even charge-neutral scalars $H_{1,2,3}$, one CP-odd charge-neutral scalar $A$, two singly-charged scalars $H^{\pm}$. 
One can secure these mass eigenstates from the unphysical fields, via the rotation matrices, as follows.
\\
For CP-even scalar sector :
\be
 \begin{pmatrix}
   H_1 \cr H_2 \cr H_3
  \end{pmatrix}
   = O_{\a}
  \begin{pmatrix}
   \rho_1 \cr \rho_2 \cr \rho_s
  \end{pmatrix}\,,
\qquad
{\rm where}, O_{\a}\,\, {\rm is\,\, a}\,\, 3\times3\,\, {\rm orthogonal\,\, matrix}. 
\ee
The hierarchy of the masses of the CP-even scalars is assumed here such that, 
$m_{H_1}\le m_{H_2} \le m_{H_3}$, 
where $m_{H_i}$ is the mass of $H_i$. 
One of these three Higgs bosons must be considered to be the SM Higgs boson. 
The scalar $H_2$ is treated here as the observed SM Higgs boson with mass at about $125$ GeV. 
The matrix $O_{\a}$ diagonalises the mass matrix (Eqn. \ref{eq:MassMatrix_N2HDM}) of the CP-even Higgs sector. 
In dark phase, either of the doublet VEVs $v_j$ or the singlet VEV $u$ can be zero. 
The rotation matrix $O_{\a}$ is simplified to the form given in the Eqns. \ref{eq:Oalpha1} and \ref{eq:Oalpha3}, respectively, for $v_2=0$ or $v_1=0$. 
The rotation matrix $O_{\a}$ is simplified again for $u=0$, and is given in the Eqn. \ref{eq:Oalpha2}. 
In the broken phase, with all the VEVs having non-zero values, the rotation matrix is in its most general form as given in the Eqn. \ref{eq:Oalpha}. 
\\
For CP-odd and charged scalar sectors :
\be
 \begin{pmatrix}
   G^0 \cr A
  \end{pmatrix}
   = O_{\b}
  \begin{pmatrix}
   \eta_1 \cr \eta_2
  \end{pmatrix}\,,\quad
 \begin{pmatrix}
   G^{\pm} \cr H^{\pm}
  \end{pmatrix}
   = O_{\b}
  \begin{pmatrix}
   \phi_1^{\pm} \cr \phi_2^{\pm}
  \end{pmatrix}\,, \qquad 
{\rm where},\,\, O_{\b}\,\, {\rm is\,\, a}\,\, 2\times2\,\, {\rm orthogonal\,\, matrix}. 
\ee
The matrix $O_{\b}$ diagonalises the mass matrix of the CP-odd as well as the charged Higgs sector. 
\\
These rotation matrices $O_{\a},\,O_{\b}$ are given in the Appendix \ref{A}. 
\\\\
Now, to focus on the couplings between the Higgs and the fermions, one needs to consider the Yukawa Lagrangian. 
In this model, the Higgs couplings to the fermions with mass $m_f$ are not as simple as that in the SM, where the coupling is $m_f/v$. 
In the two Higgs doublet model (2HDM), with the aim of averting the Flavor Changing Neutral Current (FCNC) at the tree level, four different models with diverse couplings between the Higgs and the fermions are introduced as well as probed in the literature \cite{Branco:2011iw}. 
In the Type-I Yukawa structure, the masses of the up-type quarks, down-type quarks, leptons are generated by the doublet $\Phi_2$. 
In the Type-II Yukawa structure, the mass of the up-type quarks is generated by the doublet $\Phi_2$, whereas the masses of the down-type quarks, and the leptons are generated by the doublet $\Phi_1$. 
In the Type-X Yukawa structure, the masses of the up and down type quarks are generated by the doublet $\Phi_2$, while the mass of the leptons is generated by the doublet $\Phi_1$. 
In the Type-Y Yukawa structure, the masses of the up-type quarks and leptons are generated by the doublet $\Phi_2$, in contrast the mass of the down-type quarks is generated by the doublet $\Phi_1$. 
In the N2HDM, all types of Yukawa structures have already been posited for 2HDM, will remain the same, and no more variety of Yukawa Lagrangian is introduced. 
\\
One can write the Yukawa Lagrangian as, 
\be
 {\cal{L}}_{Yukawa}^{N2HDM} = 
 - \Big[ 
 \overline{Q}_L \widetilde{\Phi}_l {\cal{Y}}_u u_R 
 \,+\, \overline{Q}_L \Phi_m {\cal{Y}}_d d_R 
 \,+\, \overline{L}_L \Phi_n {\cal{Y}}_l l_R
 \,+\, {\rm h.c.} \Big]\,,
\label{eq:Yukawa_N2HDM}
\ee
where, $Q_L,\,L_L$ represent the quark doublet and the lepton doublet, while $u_R,\,d_R,\,l_R$ represent the singlets of up-type quark, down-type quark and lepton, respectively. 
${\cal{Y}}_{u,d,l}$ are the matrices in the flavor space, and $\widetilde{\Phi}_l \equiv i\sigma_2\Phi_l^{{}^{\star}}$. 
The values of the subscripts $l,\,m,\,n$ of the unphysical scalars $\Phi$ (or $\widetilde{\Phi}$) are guided by the type of the Yukawa Lagrangian as already stated above and given in the Table \ref{tab:N2HDM}. 
Using the components of the unphysical fields, one can rewrite the Eqn. \ref{eq:Yukawa_N2HDM} in terms of the physical scalars ($\mathtt{s} =$ $\mathfrak{s}$, $A$, $H^{\pm}$, where the CP-even neutral scalars are given by $\mathfrak{s} =$ $H_1$, $H_2$, $H_3$), the fermions ($f=$ $u$, $d$, $l$), and the corresponding couplings $g_{\mathtt{s}}^f = (m_f/v)\,\xi_{\mathtt{s}}^f$. 
Clearly the Yukawa couplings in this model scaled to the Yukawa coupling in the SM is denoted by $\xi_{\mathtt{s}}^f$. 
The total Yukawa Lagrangian can thus be represented as the summation of three parts :
\be
 {\cal{L}}_{Yukawa}^{N2HDM} 
 = 
 {\cal{L}}_{Y,\mathfrak{s}}^{N2HDM} \,+\,
 {\cal{L}}_{Y,A}^{N2HDM} \,+\, 
 {\cal{L}}_{Y,H^{\pm}}^{N2HDM}
\label{eq:YukawaPart_N2HDM}
\ee
Here, the subscripts of ${\cal{L}}$ denotes the coupling of that Higgs with the fermions. Thus, the three terms represent the part of the Yukawa Lagrangian possesing $\mathfrak{s}$, $A$, $H^{\pm}$ respectively, and the pair of fermions. 
Since this paper only concentrates on the CP-even neutral scalars, only the first part of the Eqn. \ref{eq:YukawaPart_N2HDM} is given here by, 
\be
 {\cal{L}}_{Y,\mathfrak{s}}^{N2HDM} = 
 - \sum_{f=u,d,l} \frac{m_f}{v} \Big( \xi_{H_1}^f \bar{f} f H_1 \,+\, \xi_{H_2}^f \bar{f} f H_2 \,+\, \xi_{H_3}^f \bar{f} f H_3 \Big)\,, 
\label{eq:xi_N2HDM}
\ee
with $\xi_{H_{1,2,3}}^f$ listed in the Table \ref{tab:N2HDM} for the four types of Yukawa Lagrangian. 
\\
 \begin{table}
 \small
  \begin{center}
   \begin{tabular}{|| c || c | c | c | c ||}
    \hline\hline
         & Type - I & Type - II & Type - X & Type - Y \\
    \hline\hline\hline
    (l,m,n) & (2,2,2) & (2,1,1) & (2,2,1) & (2,1,2) \\ \hline\hline
    $\xi_{H_1}^u$ & $O_{\a_{12}}/\sin{\b}$ & $O_{\a_{12}}/\sin{\b}$ & $O_{\a_{12}}/\sin{\b}$ & $O_{\a_{12}}/\sin{\b}$ \\ \hline
    $\xi_{H_1}^d$ & $O_{\a_{12}}/\sin{\b}$ & $O_{\a_{11}}/\cos{\b}$ & $O_{\a_{12}}/\sin{\b}$ & $O_{\a_{11}}/\cos{\b}$ \\ \hline
    $\xi_{H_1}^l$ & $O_{\a_{12}}/\sin{\b}$ & $O_{\a_{11}}/\cos{\b}$ & $O_{\a_{11}}/\cos{\b}$ & $O_{\a_{12}}/\sin{\b}$ \\ \hline\hline
    $\xi_{H_2}^u$ & $O_{\a_{22}}/\sin{\b}$ & $O_{\a_{22}}/\sin{\b}$ & $O_{\a_{22}}/\sin{\b}$ & $O_{\a_{22}}/\sin{\b}$ \\ \hline
    $\xi_{H_2}^d$ & $O_{\a_{22}}/\sin{\b}$ & $O_{\a_{21}}/\cos{\b}$ & $O_{\a_{22}}/\sin{\b}$ & $O_{\a_{21}}/\cos{\b}$ \\ \hline
    $\xi_{H_2}^l$ & $O_{\a_{22}}/\sin{\b}$ & $O_{\a_{21}}/\cos{\b}$ & $O_{\a_{21}}/\cos{\b}$ & $O_{\a_{22}}/\sin{\b}$ \\ \hline\hline
    $\xi_{H_3}^u$ & $O_{\a_{32}}/\sin{\b}$ & $O_{\a_{32}}/\sin{\b}$ & $O_{\a_{32}}/\sin{\b}$ & $O_{\a_{32}}/\sin{\b}$ \\ \hline
    $\xi_{H_3}^d$ & $O_{\a_{32}}/\sin{\b}$ & $O_{\a_{31}}/\cos{\b}$ & $O_{\a_{32}}/\sin{\b}$ & $O_{\a_{31}}/\cos{\b}$ \\ \hline
    $\xi_{H_3}^l$ & $O_{\a_{32}}/\sin{\b}$ & $O_{\a_{31}}/\cos{\b}$ & $O_{\a_{31}}/\cos{\b}$ & $O_{\a_{32}}/\sin{\b}$ \\ \hline\hline
   \end{tabular}
   \caption{\small Yukawa couplings of the CP-even neutral Higgs bosons $H_{1,2,3}$ to the fermions $u,\,d,\,l$ in the four different models of N2HDM  scaled to Yukawa coupling of the Higgs boson to the fermions in the SM. $(O_{\a})_{p,q}$ with $p=1,2,3$, $q=1,2$ are expressed in the Eqns. \ref{eq:Oalpha}, \ref{eq:Oalpha1}, \ref{eq:Oalpha2}, \ref{eq:Oalpha3} to be used according to the general or special cases.}
  \end{center}
  \label{tab:N2HDM}
 \end{table}
\\
The kinetic part of the Lagrangian of N2HDM can be expressed in terms of the weak eigenstates $\Phi_1,\,\Phi_2,\,\chi$ as : 
\be
 {\cal{L}}_{kin}^{N2HDM} = 
 (D^{\mu} \Phi_1)^{\dag} (D_{\mu} \Phi_1) \,+\, (D^{\mu} \Phi_2)^{\dag} (D_{\mu} \Phi_2) \,+\, \frac12 (\p^{\mu} \chi) (\p_{\mu} \chi)\,, 
\label{eq:Lkin_N2HDM}
\ee
which gives the ratio ($C_{H_{1,2,3}}^{VV}$) of couplings of the CP-even neutral scalars $H_{1,2,3}$ with the gauge bosons pair $VV = WW,\,ZZ$ ($g_{H_{1,2,3}}^{VV}$) to the same coupling in the SM ($g_{H_{SM}}^{VV}$) as : 
\bea
 C_{H_1}^{VV} &=& 
 \cos{\b}\,\, O_{\a_{11}} \,+\, \sin{\b}\,\,O_{\a_{12}} \nonumber\\
 C_{H_2}^{VV} &=& 
 \cos{\b}\,\, O_{\a_{21}} \,+\, \sin{\b}\,\,O_{\a_{22}} \nonumber\\
 C_{H_3}^{VV} &=& 
 \cos{\b}\,\, O_{\a_{31}} \,+\, \sin{\b}\,\,O_{\a_{32}} 
\label{eq:HVV_N2HDM}
\eea
\\
The nine unphysical scalars can be expressed in terms of the physical scalars, Goldstone bosons, and the components $(O_\a)_{p,q},\,(O_\b)_{a,b}$, with $(p,q=1,2,3)$, $(a,b=1,2)$ of the rotation matrices $O_{\a},\,O_{\b}$ as, 
\bea
 \rho_1 &=& 
 O_{\alpha_{11}} H_1 + O_{\alpha_{21}} H_2 + O_{\alpha_{31}} H_3 \nonumber\\
 \rho_2 &=& 
 O_{\alpha_{12}} H_1 + O_{\alpha_{22}} H_2 + O_{\alpha_{32}} H_3 \nonumber\\
 \rho_s &=& 
 O_{\alpha_{13}} H_1 + O_{\alpha_{23}} H_2 + O_{\alpha_{33}} H_3 \nonumber\\
 \eta_1 &=& 
 O_{\b_{11}} G^0 + O_{\b_{21}} A \nonumber\\
 \eta_2 &=& 
 O_{\b_{12}} G^0 + O_{\b_{22}} A \nonumber\\
 \phi_1^{\pm} &=& 
 O_{\b_{11}} G^{\pm} + O_{\b_{21}} H^{\pm} \nonumber\\
 \phi_2^{\pm} &=& 
 O_{\b_{12}} G^{\pm} + O_{\b_{22}} H^{\pm}
\label{eq:unphysicalN2HDM}
\eea
\\
Probing Eqn. \ref{eq:unphysicalN2HDM} into Eqn.\ref{eq:fields}, one can write the neutral components of the doublets and the singlet as, 
\bea
 \phi_1^0 &=& 
 \frac{1}{\sqrt{2}} (v_1 + i O_{\b_{11}} G^0 + i O_{\b_{21}} A + O_{\alpha_{11}} H_1 + O_{\alpha_{21}} H_2 + O_{\alpha_{31}} H_3 ) \nonumber\\
 \phi_2^0 &=& 
 \frac{1}{\sqrt{2}} (v_2 + i O_{\b_{12}} G^0 + i O_{\b_{22}} A + O_{\alpha_{12}} H_1 + O_{\alpha_{22}} H_2 + O_{\alpha_{32}} H_3) \nonumber\\ 
 \chi &=& u + O_{\alpha_{13}} H_1 + O_{\alpha_{23}} H_2 + O_{\alpha_{33}} H_3 
\label{eq:neutralN2HDM}
\eea
\\
We assign the physical charged and neutral scalars, as,
\bea
 S_1^{+}=G^{+},\,\,S_2^{+}=H^{+},
 S_1^0=G^0,\,\,S_2^0=H_1,\,\,S_3^0=A,\,\,S_4^0=H_2,\,\,S_5^0=H_3.\,\,
\label{eq:fieldsN2HDM}
\eea
\\
We have $n_D=2$ charged scalar fields, $S_{1,2}^+$, out of which $S_1^+$ 
is the Goldstone boson, and $m=5$ real neutral scalar fields, $S_{1,2,3,4,5}^0$\,\,, out of which $S_1^0$ is the Goldstone boson, and the rest are the neutral physical scalars. 
The charged scalars $S_{1,2}^+$ are connected to the unphysical scalars $\phi_{1,2}^+$ through the matrix $\cal{U}$. 
The matrices $\cal{V}$ and $\cal{R}$ associate with the neutral components of the doublets $\Phi_{1,2}$ and that of the singlet $\chi$ to the physical neutral scalars $S_{1,2,3,4,5}^0$ respectively. 
Following the procedures described in the general section \ref{GeneralModel}, in terms of the matrices $\cal{U,\,V,\,R}$, and the newly defined scalars $S_{1,2}^+$ and $S_{1,2,3,4,5}^0$ as given in the Eqn. \ref{eq:fieldsN2HDM}, the components of the doublets $\Phi_1,\,\Phi_2$ and the singlet $\chi$ can be expressed as,
\bea
 \phi_j^{+}=\sum_{a=1}^2 {\cal{U}}_{ja}S_a^+\,\,,
 \phi_{j}^{0^{\prime}}=\sum_{b=1}^5 {\cal{V}}_{jb}S_b^0\,\,,
 \chi^{{\prime}}=\sum_{b=1}^5 {\cal{R}}_{1b}S_b^0\,\,,
{\rm with}\,\, {j}=1,2.
\label{eq:compmatN2HDM}
\eea
\\
The matrices ${\cal{U}}$, ${\cal{V}}$ and ${\cal{R}}$ are of dimensions $2\times2$, $2\times5$ and $1\times5$ respectively. 
\\
Also, using the Eqns. \ref{eq:phichiN2HDM} and \ref{eq:neutralN2HDM}, one can write the components $\phi_1^+,\,\phi_1^{0^{\prime}},\,\phi_2^+,\,\phi_2^{0^{\prime}},\,\chi^{\prime}$ in terms of the components of the mixing matrices $O_{\a},\,O_{\b}$ and the scalars defined in the Eqn. \ref{eq:fieldsN2HDM} as, 
\bea
 \phi_1^+ &=& 
 O_{\b_{11}} S_1^+ + O_{\b_{21}} S_2^+\,, \nonumber\\
 \phi_2^+ &=& 
 O_{\b_{12}} S_1^+ + O_{\b_{22}} S_2^+\,,
 \nonumber\\
 \phi_1^{0^{\prime}} &=& 
 i\, O_{\b_{11}} S_1^0 + i\, O_{\b_{21}} S_3^0 + O_{\alpha_{11}} S_2^0 + O_{\alpha_{21}} S_4^0 + O_{\alpha_{31}} S_5^0\,, \nonumber\\
 \phi_2^{0^{\prime}} &=& 
 i\, O_{\b_{12}} S_1^0 + i\, O_{\b_{22}} S_3^0 + O_{\alpha_{12}} S_2^0 + O_{\alpha_{22}} S_4^0 + O_{\alpha_{32}} S_5^0\,, \nonumber\\
 \chi^{\prime} &=& 
 O_{\alpha_{13}} S_2^0 + O_{\alpha_{23}} S_4^0 + O_{\alpha_{33}} S_5^0\,. 
\label{eq:phichiprimeN2HDM}
\eea
Therefore, comparing the Eqns. \ref{eq:phichiprimeN2HDM} with \ref{eq:compmatN2HDM} the matrices
are given by,
\bea
 {\cal{U}}=(O_{\b})^T\,, \quad
 &&{\cal{V}}=
  \begin{pmatrix}
  0 & O_{\a_{11}} & 0 & O_{\a_{21}} & O_{\a_{31}} \cr
  0 & O_{\a_{12}} & 0 & O_{\a_{22}} & O_{\a_{32}}
 \end{pmatrix}\,
 + i\,
  \begin{pmatrix}
  O_{\b_{11}} & 0 & O_{\b_{21}} & 0 & 0 \cr
  O_{\b_{12}} & 0 & O_{\b_{22}} & 0 & 0
 \end{pmatrix}\,,\n\\
 &&{\cal{R}}=
 \begin{pmatrix}
  0 & O_{\a_{13}} & 0 & O_{\a_{23}} & O_{\a_{33}} 
 \end{pmatrix}\,.
  \label{eq:UVR2}
\eea
\begin{figure}
 \begin{center}
 \includegraphics[width= 8.2cm]{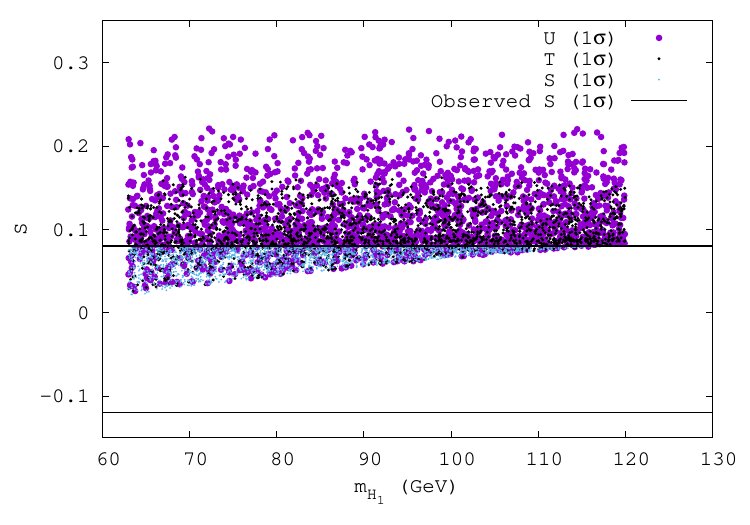} \ \
 \includegraphics[width= 8.2cm]{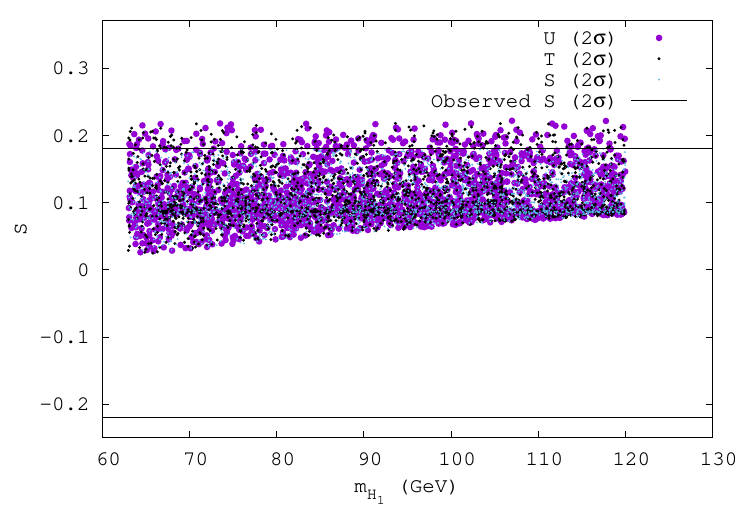}
 \end{center}
 \caption{\small The oblique parameter $\mathbb{S}$ as the function of the mass of the lightest CP-even neutral Higgs boson ($H_1$) in the model Rx2SM. The violet, black, and cyan points indicate the values of the $\mathbb{S}$ parameter when $only$ the $\mathbb{U}$, $\mathbb{T}$, and $\mathbb{S}$ parameter is constrained within the $1\sigma$ (left plot) or $2\sigma$ (right plot) region of the observed data, which is represented by the solid black line. }
 \label{fig:Rx2SM_S_m1}
 \end{figure}

\subsection{The Three Higgs Doublet Model (3HDM)}
\label{3HDM model}

The scalar sector of this model consists of three complex $SU(2)_L$ doublets with hypercharge $Y = 1$, where $j$-th doublet can be expanded concerning the component fields, as,
$ \Phi_j = \Big( \phi_j^{+}\,\,\phi_j^{0} \Big)^T $ 
 with j = 1,2,3.
\\
One can write the scalar potential as : 
\bea
 V(\Phi_1,\Phi_2,\Phi_3) &=& 
 \mu_{11}^2 (\Phi_1^{\dag} \Phi_1) \,+\, \mu_{22}^2 (\Phi_2^{\dag} \Phi_2) \,+\, \mu_{33}^2 (\Phi_3^{\dag} \Phi_3) \nonumber\\ 
 &+& \l_1 (\Phi_1^{\dag} \Phi_1)^2 \,+\, \l_2 (\Phi_2^{\dag} \Phi_2)^2 \,+\, \l_3 (\Phi_3^{\dag} \Phi_3)^2 \nonumber\\
 &+& 
 \l_4 (\Phi_1^{\dag} \Phi_1) (\Phi_2^{\dag} \Phi_2) \,+\, \l_5 (\Phi_1^{\dag} \Phi_1) (\Phi_3^{\dag} \Phi_3) \,+\, \l_6 (\Phi_2^{\dag} \Phi_2) (\Phi_3^{\dag} \Phi_3) \nonumber\\
 &+& 
 \l_7 (\Phi_1^{\dag} \Phi_2) (\Phi_2^{\dag} \Phi_1) \,+\, \l_8 (\Phi_1^{\dag} \Phi_3) (\Phi_3^{\dag} \Phi_1) \,+\, \l_9 (\Phi_2^{\dag} \Phi_3) (\Phi_3^{\dag} \Phi_2) \nonumber\\
 &+& 
 \l_{10} [(\Phi_1^{\dag} \Phi_2)(\Phi_1^{\dag} \Phi_3) + (\Phi_2^{\dag} \Phi_1)(\Phi_3^{\dag} \Phi_1)] \,+\, \l_{11} [(\Phi_1^{\dag} \Phi_2)(\Phi_3^{\dag} \Phi_2) + (\Phi_2^{\dag} \Phi_1)(\Phi_2^{\dag} \Phi_3)] \nonumber\\
 &+& 
 \l_{12} [(\Phi_1^{\dag} \Phi_3)(\Phi_2^{\dag} \Phi_3) + (\Phi_3^{\dag} \Phi_1)(\Phi_3^{\dag} \Phi_2)]
\label{eq:pot_3HDM}
\eea
Here, $\mu_{11,22,33}$ are of dimension $2$, and $\l_{1-12}$ are dimensionless coefficients. 
The parameters $\mu_{11}$, $\mu_{22}$, $\mu_{33}$, $\l_1$, $\l_2$, $\l_3$, $\l_4$, $\l_6$, $\l_7$, $\l_8$, $\l_9$ are real while $\l_{10}$, $\l_{11}$, $\l_{12}$ can be complex, but this paper assumes all these parameters to be real \cite{Coleppa:2025qst}. 
This scalar potential is symmetric under the $Z_3$, where the Higgs fields transform as $\Phi_1 \rightarrow \omega \Phi_1$, $\Phi_2 \rightarrow \omega^2 \Phi_2$, $\Phi_3 \rightarrow \Phi_3$. 
Here, the cube roots of unity $\omega$ is given by, $\omega = e^{2\pi i/3}$. 
\\
After SSB, the VEV for $j$-th neutral field is $<\phi_j^0>=v_j/\sqrt{2}$, such that, the total electroweak VEV, $v$, can be expressed as,
$ v^2=\sum_{j=1}^3 v_j^2 = (246 \rm\,  GeV)^2$.
\\
After expanding the neutral fields about their VEVs, 
\bea
 \phi_j^0=\frac{1}{\sqrt{2}}(v_j+\phi_j^{0^{\prime}})\,, \qquad {\rm with}\,\,
 \phi_j^{0^{\prime}}= \rho_j+i\,\eta_j\,.
\label{eq:phi3HDM} 
\eea 
The three doublet VEVs can be expressed in terms of two mixing angles $\b_1$, $\b_2$, such that, $v_1=v \cos{\b_1} \cos{\b_2}$, $v_2=v \sin{\b_1} \cos{\b_2}$ and $v_3=v\sin{\b_2}$, leading to the ratio of the doublet VEVs 
\be
\frac{v_2}{v_1} = \tan{\b_1}\,, \qquad
\frac{v_3}{\sqrt{v_1^2+v_2^2}} = \tan{\b_2}\,.
\label{eq:tanb_3HDM} 
\ee
\\
The VEVs help to realize different phases of this model. 
If any of the doublet VEVs $v_j$ is zero, the corresponding real part of the neutral unphysical field $\rho_j$ of the doublet $\Phi_j$ is the dark matter candidate, provided there is some symmetry in the potential, and the model is in dark phase. 
This dark matter candidate does not mix with other unphysical scalars and has no couplings to the SM particles. 
For $v_1,\,v_2,\,v_3 \ne0$, all the three unphysical fields $\rho_1,\,\rho_2,\,\rho_3$ mix to provide the physical fields $H_{1,2,3}$ resulting the broken phase. 
This paper does not consider these phases further, as those are out of scope of this discussion. 
\\
The minimization conditions of the scalar potential \ref{eq:pot_3HDM} eliminate the mass terms as :
\bea
 \mu_{11}^2 &=& 
 - \l_1 v_1^2 \,-\, \frac{(\l_4+\l_7)}{2} v_2^2 \,-\, \frac{(\l_5+\l_8)}{2} v_3^2 \,-\, \l_{10} v_2 v_3 \,-\, \frac{\l_{11}}{2} \frac{v_2^2v_3}{v_1} \,-\, \frac{\l_{12}}{2} \frac{v_2 v_3^2}{v_1} \nonumber\\
 \mu_{22}^2 &=& 
 - \l_2 v_2^2 \,-\, \frac{(\l_4+\l_7)}{2} v_1^2 \,-\, \frac{(\l_6+\l_9)}{2} v_3^2 \,-\, \frac{\l_{10}}{2} \frac{v_1^2 v_3}{v_2} \,-\, \l_{11} v_1 v_3 \,-\, \frac{\l_{12}}{2} \frac{v_1 v_3^2}{v_2} \nonumber\\
 \mu_{33}^2 &=& 
 - \l_3 v_3^2 \,-\, \frac{(\l_5+\l_8)}{2} v_1^2 \,-\, \frac{(\l_6+\l_9)}{2} v_2^2 \,-\, \frac{\l_{10}}{2} \frac{v_1^2 v_2}{v_3} \,-\, \frac{\l_{11}}{2} \frac{v_1 v_2^2}{v_3} \,-\, \l_{12} v_1 v_2 
\label{eq:eliminate_3HDM}
\eea
This paper only pays interest in the neutral CP-even scalars $\rho_1,\,\rho_2,\,\rho_3$ and hence only the corresponding mass-matrix is given here, as : 
\bea
 {\cal{M}}_{CP-even}^2 = 
 \begin{pmatrix}
  2\l_1 v_1^2 - (\l_{11} v_2 + \l_{12} v_3)\frac{v_2 v_3}{2v_1} & ({\cal{M}}_{CP-even,3HDM}^2)_{12} & ({\cal{M}}_{CP-even,3HDM}^2)_{13} \cr
  ({\cal{M}}_{CP-even,3HDM}^2)_{12} & 2\l_2 v_2^2 - (\l_{10} v_1 + \l_{12} v_3)\frac{v_1 v_3}{2v_2} & ({\cal{M}}_{CP-even,3HDM}^2)_{23} \cr
  ({\cal{M}}_{CP-even,3HDM}^2)_{13} & ({\cal{M}}_{CP-even,3HDM}^2)_{23} & 2\l_3 v_3^2 - (\l_{10} v_1 + \l_{11} v_2)\frac{v_1 v_2}{2v_3} 
 \end{pmatrix}\,,\qquad
\label{eq:MassMatrix_3HDM}
\eea
with 
\bea
 ({\cal{M}}_{CP-even,3HDM}^2)_{12} &=& 
 (\l_4+\l_7) v_1 v_2 + \l_{10} v_1 v_3 + \l_{11} v_2 v_3 + \frac{\l_{12}}{2} v_3^2 \nonumber\\
 ({\cal{M}}_{CP-even,3HDM}^2)_{13} &=& 
 (\l_5+\l_8) v_1 v_3 + \l_{10} v_1 v_2 + \frac{\l_{11}}{2} v_2^2 + \l_{12} v_2 v_3 \nonumber\\
 ({\cal{M}}_{CP-even,3HDM}^2)_{23} &=& 
 (\l_6+\l_9) v_2 v_3 + \frac{\l_{10}}{2} v_1^2 + \l_{11} v_1 v_2 + \l_{12} v_1 v_3
\label{eq:Comp_MassMatrix_3HDM} 
\eea
\\
This model contains twelve scalars, out of which three are Goldstone bosons $G^{0,\pm}$. 
3HDM also contains three CP-even charge-neutral scalars $H_{1,2,3}$, two CP-odd charge-neutral scalars $A_{1,2}$, and four singly-charged scalars $H_{1,2}^{\pm}$. 
One can secure these mass eigenstates from the unphysical fields,
via the $3\times3$ orthogonal rotation matrices ($O_{\a}$, $O_{\b\gamma_{1,2}}$), for CP-even, CP-odd and charged scalar sector, respectively, as follows.\\ 
\bea
 \begin{pmatrix}
  H_1 \cr H_2 \cr H_3
 \end{pmatrix}
  = O_{\a}
 \begin{pmatrix}
  \rho_1 \cr \rho_2 \cr \rho_3
 \end{pmatrix}\,,\quad
 \begin{pmatrix}
  G^{0} \cr A_1 \cr A_2
 \end{pmatrix}
  = O_{\b\gamma_1}
 \begin{pmatrix}
  \eta_1 \cr \eta_2 \cr \eta_3
 \end{pmatrix},\quad
 \begin{pmatrix}
  G^{\pm} \cr H_1^{\pm} \cr H_2^{\pm}
 \end{pmatrix}
  = O_{\b\gamma_2}
 \begin{pmatrix}
  \phi_1^{\pm} \cr \phi_2^{\pm} \cr \phi_3^{\pm}
 \end{pmatrix}\,.
\eea
The hierarchy of the masses of the CP-even scalars is assumed here such that, 
$m_{H_1}\le m_{H_2} \le m_{H_3}$, 
where $m_{H_i}$ is the mass of $H_i$. 
One of these three Higgs bosons must be considered to be the SM Higgs boson. 
The scalar $H_2$ is treated here as the observed SM Higgs boson with mass at about $125$ GeV. 
The matrix $O_{\a}$ diagonalises the mass matrix of the CP-even Higgs sector. 
The matrices $O_{\b\gamma_1}$ and $O_{\b\gamma_2}$ diagonalise the mass matrices of the CP-odd and the charged Higgs sector, respectively. 
The forms of these rotation matrices are given in the Appendix 
\ref{A}. 
In the dark phase, the rotation matrix $O_{\a}$ is simplified to the form given in the Eqns. \ref{eq:Oalpha1}, \ref{eq:Oalpha2} and \ref{eq:Oalpha3}, respectively, for $v_2=0$ or $v_3=0$ or $v_1=0$. 
In the broken phase, where all the doublet VEVs possess non-zero values, the rotation matrix is in its most general form as given in the Eqn. \ref{eq:Oalpha}. 
\\\\
Now, to focus on the couplings between the Higgs and the fermions, one needs to consider the Yukawa Lagrangian. 
In this model, the Higgs couplings to the fermions with mass $m_f$ are not as simple as that in the SM, where the coupling is $m_f/v$. 
In this model, with the intention of avoiding the Flavor Changing Neutral Current (FCNC) at the tree level, five different models with different couplings between the Higgs and the fermions are introduced. 
The first four models (type - I, II, X, Y) are same as the 2HDM or N2HDM, while the fifth model (type - Z) is completely new. 
In the Type-I Yukawa structure, the masses of the up-type quarks, down-type quarks, leptons are generated by the doublet $\Phi_2$. 
In the Type-II Yukawa structure, the mass of the up-type quarks is generated by the doublet $\Phi_2$, whereas the masses of the down-type quarks, and the leptons are generated by the doublet $\Phi_1$. 
In the Type-X Yukawa structure, the masses of the up and down type quarks are generated by the doublet $\Phi_2$, while the mass of the leptons is generated by the doublet $\Phi_1$. 
In the Type-Y Yukawa structure, the masses of the up-type quarks and leptons are generated by the doublet $\Phi_2$, in contrast the mass of the down-type quarks is generated by the doublet $\Phi_1$. 
In the 3HDM, along with types of Yukawa structures have already been posited for 2HDM or N2HDM, one more variety of Yukawa Lagrangian is introduced. 
In the type-Z Yukawa structure, the masses of the up-type quarks, down-type quarks, and the leptons are generated by the doublets $\Phi_2$, $\Phi_1$, and $\Phi_3$ respectively. 
\begin{figure}
 \begin{center}
 \includegraphics[width= 8.2cm]{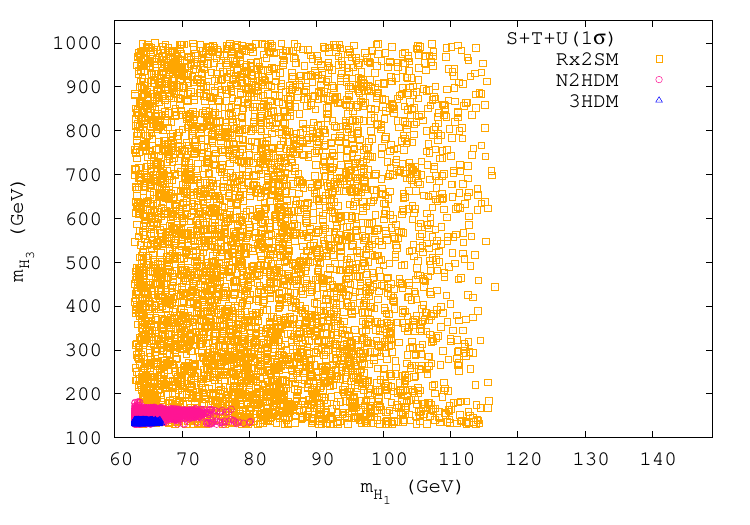} \ \
 \includegraphics[width= 8.2cm]{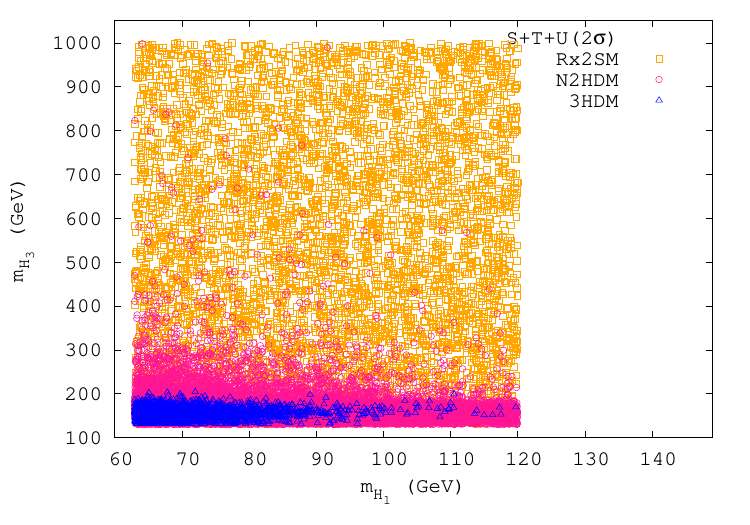}
 \end{center}
 \caption{\small The mass of the heaviest CP-even neutral Higgs boson ($H_3$) as the function of the mass of the lightest CP-even neutral Higgs boson ($H_1$), while setting $m_{H_2}\approx 125$ GeV. The squares (orange points), the circles (pink points), and the triangles (blue points) indicate the allowed region in the mass plane for the Rx2SM, N2HDM, and 3HDM models respectively, when all of the $\mathbb{S}$, $\mathbb{T}$, $\mathbb{U}$ parmeters are either within $1\sigma$ (left plot) or $2\sigma$ (right plot). }
 \label{fig:STU_m1m3}
 \end{figure}
\\
 \begin{table}
 \small
  \begin{center}
   \begin{tabular}{|| c || c | c | c | c | c ||}
    \hline\hline
         & Type - I & Type - II & Type - X & Type - Y & Type - Z \\
    \hline\hline\hline
    (l,m,n) & (2,2,2) & (2,1,1) & (2,2,1) & (2,1,2) & (2,1,3) \\ \hline\hline
    $\xi_{H_1}^u$ & $O_{\a_{12}}/\sin{\b_1}\cos{\b_2}$ & $O_{\a_{12}}/\sin{\b_1}\cos{\b_2}$ & $O_{\a_{12}}/\sin{\b_1}\cos{\b_2}$ & $O_{\a_{12}}/\sin{\b_1}\cos{\b_2}$ & $O_{\a_{12}}/\sin{\b_1}\cos{\b_2}$ \\ \hline
    $\xi_{H_1}^d$ & $O_{\a_{12}}/\sin{\b_1}\cos{\b_2}$ & $O_{\a_{11}}/\cos{\b_1}\cos{\b_2}$ & $O_{\a_{12}}/\sin{\b_1}\cos{\b_2}$ & $O_{\a_{11}}/\cos{\b_1}\cos{\b_2}$ & $O_{\a_{11}}/\cos{\b_1}\cos{\b_2}$ \\ \hline
    $\xi_{H_1}^l$ & $O_{\a_{12}}/\sin{\b_1}\cos{\b_2}$ & $O_{\a_{11}}/\cos{\b_1}\cos{\b_2}$ & $O_{\a_{11}}/\cos{\b_1}\cos{\b_2}$ & $O_{\a_{12}}/\sin{\b_1}\cos{\b_2}$ & $O_{\a_{13}}/\sin{\b_2}$ \\ \hline\hline
    $\xi_{H_2}^u$ & $O_{\a_{22}}/\sin{\b_1}\cos{\b_2}$ & $O_{\a_{22}}/\sin{\b_1}\cos{\b_2}$ & $O_{\a_{22}}/\sin{\b_1}\cos{\b_2}$ & $O_{\a_{22}}/\sin{\b_1}\cos{\b_2}$ & $O_{\a_{22}}/\sin{\b_1}\cos{\b_2}$ \\ \hline
    $\xi_{H_2}^d$ & $O_{\a_{22}}/\sin{\b_1}\cos{\b_2}$ & $O_{\a_{21}}/\cos{\b_1}\cos{\b_2}$ & $O_{\a_{22}}/\sin{\b_1}\cos{\b_2}$ & $O_{\a_{21}}/\cos{\b_1}\cos{\b_2}$ & $O_{\a_{21}}/\cos{\b_1}\cos{\b_2}$ \\ \hline
    $\xi_{H_2}^l$ & $O_{\a_{22}}/\sin{\b_1}\cos{\b_2}$ & $O_{\a_{21}}/\cos{\b_1}\cos{\b_2}$ & $O_{\a_{21}}/\cos{\b_1}\cos{\b_2}$ & $O_{\a_{22}}/\sin{\b_1}\cos{\b_2}$ & $O_{\a_{23}}/\sin{\b_2}$ \\ \hline\hline
    $\xi_{H_3}^u$ & $O_{\a_{32}}/\sin{\b_1}\cos{\b_2}$ & $O_{\a_{32}}/\sin{\b_1}\cos{\b_2}$ & $O_{\a_{32}}/\sin{\b_1}\cos{\b_2}$ & $O_{\a_{32}}/\sin{\b_1}\cos{\b_2}$ & $O_{\a_{32}}/\sin{\b_1}\cos{\b_2}$ \\ \hline
    $\xi_{H_3}^d$ & $O_{\a_{32}}/\sin{\b_1}\cos{\b_2}$ & $O_{\a_{31}}/\cos{\b_1}\cos{\b_2}$ & $O_{\a_{32}}/\sin{\b_1}\cos{\b_2}$ & $O_{\a_{31}}/\cos{\b_1}\cos{\b_2}$ & $O_{\a_{31}}/\cos{\b_1}\cos{\b_2}$ \\ \hline
    $\xi_{H_3}^l$ & $O_{\a_{32}}/\sin{\b_1}\cos{\b_2}$ & $O_{\a_{31}}/\cos{\b_1}\cos{\b_2}$ & $O_{\a_{31}}/\cos{\b_1}\cos{\b_2}$ & $O_{\a_{32}}/\sin{\b_1}\cos{\b_2}$ & $O_{\a_{33}}/\sin{\b_2}$ \\ \hline\hline
   \end{tabular}
   \caption{\small Yukawa couplings of the CP-even neutral Higgs bosons $H_{1,2,3}$ to the fermions $u,\,d,\,l$ in the five different models of 3HDM  scaled to Yukawa coupling of the Higgs boson to the fermions in the SM. $(O_{\a})_{p,q}$ with $p,q=1,2,3$ are expressed in the Eqns. \ref{eq:Oalpha}, \ref{eq:Oalpha1}, \ref{eq:Oalpha2}, \ref{eq:Oalpha3} to be used according to the general or special cases.}
  \end{center}
  \label{tab:3HDM}
 \end{table}
\\
One can write the Yukawa Lagrangian as, 
\be
 {\cal{L}}_{Yukawa}^{3HDM} = 
 - \Big[ 
 \overline{Q}_L \widetilde{\Phi}_l {\cal{Y}}_u u_R 
 \,+\, \overline{Q}_L \Phi_m {\cal{Y}}_d d_R 
 \,+\, \overline{L}_L \Phi_n {\cal{Y}}_l l_R
 \,+\, {\rm h.c.} \Big]\,,
\label{eq:Yukawa_3HDM}
\ee
where, $Q_L,\,L_L$ represent the quark doublet and the lepton doublet, while $u_R,\,d_R,\,l_R$ represent the singlets of up-type quark, down-type quark and lepton, respectively. 
${\cal{Y}}_{u,d,l}$ are the matrices in the flavor space, and $\widetilde{\Phi}_l \equiv i\sigma_2\Phi_l^{{}^{\star}}$. 
The values of the subscripts $l,\,m,\,n$ of the unphysical scalars $\Phi$ (or $\widetilde{\Phi}$) are guided by the type of the Yukawa Lagrangian as already stated above and given in the Table \ref{tab:3HDM}. 
Using the components of the unphysical fields, one can rewrite the Eqn. \ref{eq:Yukawa_3HDM} in terms of the physical scalars ($\mathtt{s} =$ $\mathfrak{s}$, $A_1,\,A_2$, $H_1^{\pm},\,H_2^{\pm}$, where the CP-even neutral scalars are given by $\mathfrak{s} =$ $H_1$, $H_2$, $H_3$), the fermions ($f=$ $u$, $d$, $l$), and the corresponding couplings $g_{\mathtt{s}}^f = (m_f/v)\,\xi_{\mathtt{s}}^f$. 
Clearly the Yukawa couplings in this model scaled to the Yukawa coupling in the SM is denoted by $\xi_{\mathtt{s}}^f$. 
The total Yukawa Lagrangian can thus be represented as the summation of three parts :
\be
 {\cal{L}}_{Yukawa}^{3HDM} 
 = 
 {\cal{L}}_{Y,\mathfrak{s}}^{3HDM} \,+\,
 {\cal{L}}_{Y,A_{1,2}}^{3HDM} \,+\, 
 {\cal{L}}_{Y,H_{1,2}^{\pm}}^{3HDM}
\label{eq:YukawaPart_3HDM}
\ee
Here, the subscripts of ${\cal{L}}$ denotes the coupling of that Higgs with the fermions. 
Thus, the three terms represent the part of the Yukawa Lagrangian possesing $\mathfrak{s}$, $A_{1,2}$, $H_{1,2}^{\pm}$ respectively, and the pair of fermions. 
Since this paper only concentrates on the CP-even neutral scalars, only the first part of the Eqn. \ref{eq:YukawaPart_3HDM} is given here by, 
\be
 {\cal{L}}_{Y,\mathfrak{s}}^{3HDM} = 
 - \sum_{f=u,d,l} \frac{m_f}{v} \Big( \xi_{H_1}^f \bar{f} f H_1 \,+\, \xi_{H_2}^f \bar{f} f H_2 \,+\, \xi_{H_3}^f \bar{f} f H_3 \Big)\,, 
\label{eq:xi_3HDM}
\ee
with $\xi_{H_{1,2,3}}^f$ listed in the Table \ref{tab:3HDM} for the five types of Yukawa Lagrangian. 
\\\\
The kinetic part of the Lagrangian of 3HDM can be expressed in terms of the weak eigenstates $\Phi_1,\,\Phi_2,\,\Phi_3$ as : 
\be
 {\cal{L}}_{kin}^{3HDM} = 
 (D^{\mu} \Phi_1)^{\dag} (D_{\mu} \Phi_1) \,+\, (D^{\mu} \Phi_2)^{\dag} (D_{\mu} \Phi_2) \,+\, (D^{\mu} \Phi_3)^{\dag} (D_{\mu} \Phi_3)\,, 
\label{eq:Lkin_3HDM}
\ee
which gives the ratio ($C_{H_{1,2,3}}^{VV}$) of couplings of the CP-even neutral scalars $H_{1,2,3}$ with the gauge bosons pair $VV = WW,\,ZZ$ ($g_{H_{1,2,3}}^{VV}$) to the same coupling in the SM ($g_{H_{SM}}^{VV}$) as : 
\bea
 C_{H_1}^{VV} &=& 
 \cos{\b_1}\,\,\cos{\b_2}\,\, O_{\a_{11}} \,+\, \sin{\b_1}\,\,\cos{\b_2}\,\,O_{\a_{12}} \,+\, \sin{\b_2}\,\,O_{\a_{13}} \nonumber\\
 C_{H_2}^{VV} &=& 
 \cos{\b_1}\,\,\cos{\b_2}\,\, O_{\a_{21}} \,+\, \sin{\b_1}\,\,\cos{\b_2}\,\,O_{\a_{22}} \,+\, \sin{\b_2}\,\,O_{\a_{23}} \nonumber\\
 C_{H_3}^{VV} &=& 
 \cos{\b_1}\,\,\cos{\b_2}\,\, O_{\a_{31}} \,+\, \sin{\b_1}\,\,\cos{\b_2}\,\,O_{\a_{32}} \,+\, \sin{\b_2}\,\,O_{\a_{33}} 
\label{eq:HVV_3HDM}
\eea
\\
The twelve unphysical scalars can be expressed in terms of the physical scalars, Goldstone bosons, and the components $(O_{\a})_{p,q},\,(O_{{\b\gamma}_i})_{p,q}$, with $(p,q=1,2,3)$ of the rotation matrices $O_{\a},\,O_{\b\gamma_1},\,O_{\b\gamma_2}$ as, 
\bea
 \rho_1 &=& 
 O_{\alpha_{11}} H_1 + O_{\alpha_{21}} H_2 + O_{\alpha_{31}} H_3 \nonumber\\
 \rho_2 &=& 
 O_{\alpha_{12}} H_1 + O_{\alpha_{22}} H_2 + O_{\alpha_{32}} H_3 \nonumber\\
 \rho_3 &=& 
 O_{\alpha_{13}} H_1 + O_{\alpha_{23}} H_2 + O_{\alpha_{33}} H_3 \nonumber\\
 \eta_1 &=& 
 O_{{\b\gamma_1}_{11}} G^0 + O_{{\b\gamma_1}_{21}} A_1 + O_{{\b\gamma_1}_{31}} A_2 \nonumber\\
 \eta_2 &=& 
 O_{{\b\gamma_1}_{12}} G^0 + O_{{\b\gamma_1}_{22}} A_1 + O_{{\b\gamma_1}_{32}} A_2 \nonumber\\
 \eta_3 &=& 
 O_{{\b\gamma_1}_{13}} G^0 + O_{{\b\gamma_1}_{23}} A_1 + O_{{\b\gamma_1}_{33}} A_2 \nonumber\\
 \phi_1^{\pm} &=& 
 O_{{\b\gamma_2}_{11}} G^{\pm} + O_{{\b\gamma_2}_{21}} H_1^{\pm} + O_{{\b\gamma_2}_{31}} H_2^{\pm} \nonumber\\
 \phi_2^{\pm} &=& 
 O_{{\b\gamma_2}_{12}} G^{\pm} + O_{{\b\gamma_2}_{22}} H_1^{\pm} + O_{{\b\gamma_2}_{32}} H_2^{\pm} \nonumber\\
 \phi_3^{\pm} &=& 
 O_{{\b\gamma_2}_{13}} G^{\pm} + O_{{\b\gamma_2}_{23}} H_1^{\pm} + O_{{\b\gamma_2}_{33}} H_2^{\pm}\,. 
\label{eq:unphysical3HDM}
\eea
\\
Next discussion will be on the couplings of the CP-even neutral scalars $H_{1,2,3}$ with the SM particles, when all the VEVs are non-zero, $i.e.$, the phase is broken as mentioned earlier. 
\\
Probing Eqn. \ref{eq:unphysical3HDM} into Eqn.\ref{eq:fields}, one can write the neutral components of the doublets as, 
\bea
 \phi_1^0 &=& 
 \frac{1}{\sqrt{2}} (v_1 + i O_{{\b\gamma_1}_{11}} G^0 + i O_{{\b\gamma_1}_{21}} A_1 + i O_{{\b\gamma_1}_{31}} A_2 + O_{\alpha_{11}} H_1 + O_{\alpha_{21}} H_2 + O_{\alpha_{31}} H_3) \nonumber \\
 \phi_2^0 &=& 
 \frac{1}{\sqrt{2}} (v_2 + i O_{{\b\gamma_1}_{12}} G^0 + i O_{{\b\gamma_1}_{22}} A_1 + i O_{{\b\gamma_1}_{32}} A_2 + O_{\alpha_{12}} H_1 + O_{\alpha_{22}} H_2 + O_{\alpha_{32}} H_3) \nonumber \\
 \phi_3^0 &=& 
 \frac{1}{\sqrt{2}} (v_1 + i O_{{\b\gamma_1}_{13}} G^0 + i O_{{\b\gamma_1}_{23}} A_1 + i O_{{\b\gamma_1}_{33}} A_2 + O_{\alpha_{13}} H_1 + O_{\alpha_{23}} H_2 + O_{\alpha_{33}} H_3) \,.
\label{eq:neutral3HDM}
\eea
\begin{figure}
 \begin{center}
 \includegraphics[width= 8.2cm]{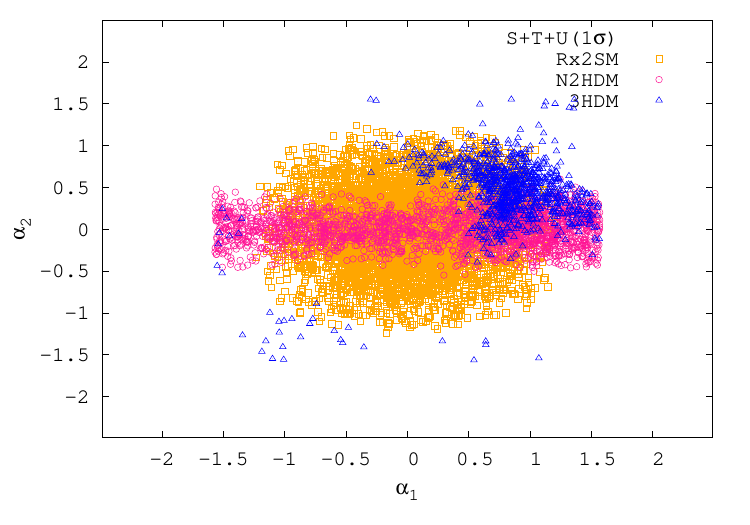} \ \
 \includegraphics[width= 8.2cm]{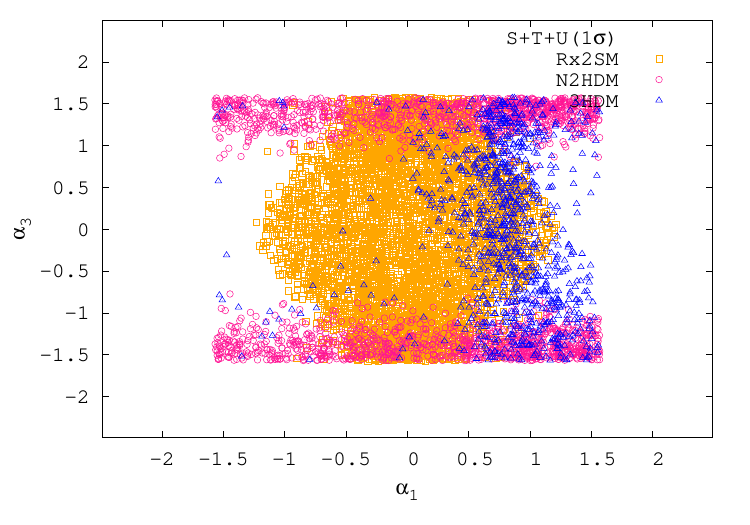}
 \end{center}
 \caption{\small The mixing angles $\a_{1,2,3}$ between the CP-even neutral Higgs bosons ($H_{1,2,3}$), while setting $m_{H_2}\approx 125$ GeV. The squares (orange points), the circles (pink points), and the triangles (blue points) indicate the allowed region for the Rx2SM, N2HDM, and 3HDM models respectively, when all of the $\mathbb{S}$, $\mathbb{T}$, $\mathbb{U}$ parmeters are within $1\sigma$. }
 \label{fig:STU_angle}
 \end{figure}
\\
We assign the physical charged and neutral scalars as,
\be
 S_1^{+}=G^{+},\,S_2^{+}=H_1^{+},\,S_3^{+}=H_2^{+},\,
 S_1^0=G^0,\,\,S_2^0=H_1,\,S_3^0=A_1,\,S_4^0=H_2,\,
 S_5^0=A_2,\,S_6^0=H_3.
\label{eq:fields3HDM} 
\ee
\\
We have $n_D=3$ charged scalar fields, $S_{1,2,3}^+$, out of which $S_1^+$ is the Goldstone boson, and $m=6$ real neutral scalar fields $S_{1,2,3,4,5,6}^0$\,\,, out of which $S_1^0$ is the Goldstone boson, and the rest are the neutral physical scalars. 
The charged scalars $S_{1,2,3}^+$ are connected to the unphysical scalars $\phi_{1,2,3}^+$ through the matrix ${\cal{U}}$. 
The matrix ${\cal{V}}$ associates with the neutral components of the doublets $\Phi_{1,2,3}$ to the physical neutral scalars $S_{1,2,3,4,5,6}^0$\,\,. 
Following the procedures described in the general section \ref{GeneralModel}, in terms of the matrices $\cal{U,\,V}$, and the newly defined scalars $S_{1,2,3}^+$ and $S_{1,2,3,4,5,6}^0$ as given in the Eqn. \ref{eq:fields3HDM}, the components of the doublets $\Phi_1,\,\Phi_2,\,\Phi_3$ can be expressed as,
\be
 \phi_j^{+}=\sum_{a=1}^3 {\cal{U}}_{ja}S_a^+,\quad
 \phi_j^{0^{\prime}}=\sum_{b=1}^6 {\cal{V}}_{jb}S_b^0\,, \quad
 {\rm with}\,\, j=1,2,3.
\label{eq:compmat3HDM} 
\ee
\\
The matrices ${\cal{U}}$ and ${\cal{V}}$ are of dimensions $3\times3$ and $3\times6$ respectively. 
\\
Also, using the Eqns. \ref{eq:phi3HDM} and \ref{eq:neutral3HDM}, one can write the components $\phi_1^+,\,\phi_1^0,\,\phi_2^+,\,\phi_2^0,\,\phi_3^+,\,\phi_3^0$ in terms of the components of the mixing matrices $O_{\a},\,O_{{\b\gamma}_{1,2}}$, and the scalars defined in the Eqn. \ref{eq:fields3HDM} as, 
\bea
 \phi_1^+ &=& 
  O_{{\b\gamma_2}_{11}} S_1^+ + O_{{\b\gamma_2}_{21}} S_2^+ + O_{{\b\gamma_2}_{31}} S_3^+ \,, \nonumber\\
 \phi_2^+ &=& 
  O_{{\b\gamma_2}_{12}} S_1^+ + O_{{\b\gamma_2}_{22}} S_2^{\pm} + O_{{\b\gamma_2}_{32}} S_3^+ \,, \nonumber\\
 \phi_3^+ &=& 
  O_{{\b\gamma_2}_{13}} S_1^+ + O_{{\b\gamma_2}_{23}} S_2^+ + O_{{\b\gamma_2}_{33}} S_3^+ \,, \nonumber\\
 \phi_1^{0^{\prime}} &=& 
 i\, O_{{\b\gamma_1}_{11}} S_1^0 + i\, O_{{\b\gamma_1}_{21}} S_3^0 + i\, O_{{\b\gamma_1}_{31}} S_5^0 + O_{\alpha_{11}} S_2^0 + O_{\alpha_{21}} S_4^0 + O_{\alpha_{31}} S_6^0 \,, \nonumber \\
 \phi_2^{0^{\prime}} &=& 
 i\, O_{{\b\gamma_1}_{12}} S_1^0 + i\, O_{{\b\gamma_1}_{22}} S_3^0 + i\, O_{{\b\gamma_1}_{32}} S_5^0 + O_{\alpha_{12}} S_2^0 + O_{\alpha_{22}} S_4^0 + O_{\alpha_{32}} S_6^0 \,, \nonumber \\
 \phi_3^{0^{\prime}} &=& 
 i\, O_{{\b\gamma_1}_{13}} S_1^0 + i\, O_{{\b\gamma_1}_{23}} S_3^0 + i\, O_{{\b\gamma_1}_{33}} S_5^0 + O_{\alpha_{13}} S_2^0 + O_{\alpha_{23}} S_4^0 + O_{\alpha_{33}} S_6^0 \,. 
\label{eq:phiprime3HDM}
\eea
Therefore, comparing the Eqns. \ref{eq:phiprime3HDM} with \ref{eq:compmat3HDM} the matrices
are given by,
\be
 {\cal{U}}=(O_{\b\gamma_2})^T\,,
 {\cal{V}}=
 \begin{pmatrix}
  0 & O_{\a_{11}} & 0 & O_{\a_{21}} & 0 & O_{\a_{31}}\cr
  0 & O_{\a_{12}} & 0 & O_{\a_{22}} & 0 & O_{\a_{32}}\cr
  0 & O_{\a_{13}} & 0 & O_{\a_{23}} & 0 & O_{\a_{33}}
 \end{pmatrix}\,
 + i 
  \begin{pmatrix}
  (O_{{\b\gamma_1})_{11}} & 0 & (O_{{\b\gamma_1})_{21}} 
  & 0 & (O_{{\b\gamma_1})_{31}} & 0\cr
  (O_{{\b\gamma_1})_{12}} & 0 & (O_{{\b\gamma_1})_{22}} 
  & 0 & (O_{{\b\gamma_1})_{32}} & 0\cr
  (O_{{\b\gamma_1})_{13}} & 0 & (O_{{\b\gamma_1})_{23}} 
  & 0 & (O_{{\b\gamma_1})_{33}} & 0
 \end{pmatrix}\,.
  \label{eq:UVR3}
\ee 
The matrix ${\cal{R}}$ does not exist in the 3HDM as it does not contain any $SU(2)_L$ singlet scalar. 

\section{Results}
\label{Results}

\numberwithin{equation}{section}

Here, we enlist the expressions of the oblique parameters concerning the components of the rotation matrices of the scalar sectors of the three models.
For this, we also need the definitions of
some well known functions \cite{Veltman:1977kh, Grimus:2007if, 
Grimus:2008nb}, as,

\bea
 F(x,y)&=&
 \begin{cases}
    \frac{x+y}{2}-\frac{xy}{x-y}\ln\frac{x}{y}\,\hspace{1cm} \rm{for}\,\,
    x\neq y\,,\n\\
    0\,\hspace{3.27cm} \rm{for}\,\, x=y\,,
 \end{cases}\\
 G(x,y,z)&=&-\frac{16}{3}+\frac{5(x+y)}{z}-\frac{2(x-y)^2}{z^2}
 +\frac{3}{z}\left(\frac{x^2+y^2}{x-y}-\frac{x^2-y^2}{z}
 +\frac{(x-y)^3}{3z^2}\ln\frac{x}{y}\right)\,\n\\
 &&+\frac{(z^2-2z(x+y)+(x-y)^2)}{z^3}
 f\left((x+y-z),(z^2-2z(x+y)+(x-y)^2)\right)\,,\n\\
 H(x,y,z)&=&2-\frac{9(x+y)}{z}+\frac{6(x-y)^2}{z^2}\,
 +\frac{3}{z}\left(-\frac{x^2+y^2}{x-y}+2\frac{x^2-y^2}{z}
 -\frac{(x-y)^3}{z^2}\right)\ln\frac{x}{y}\,\n\\
 &&+\left(x+y-\frac{(x-y)^2}{z}\right)
 \frac{3f\left((x+y-z),(z^2-2z(x+y)+(x-y)^2)\right)}{z^2}
 \,,\n\\
 \hat{G}(x,y)&=&-\frac{79}{3}+9\frac{x}{y}-2\frac{x^2}{y^2}+ 
 \left( -10+18\frac{x}{y} -6\frac{x^2}{y^2}+\frac{x^3}{y^3}
 -9\frac{x+y}{x-y}\right)\ln\frac{x}{y}\,\n\\
 &&+\left(12-4\frac{x}{y}+\frac{x^2}{y^2}\right)\frac{f(x,x^2-4xy)}{y}\,,\n\\
 \hat{H}(x,y)&=& 47-21\frac{x}{y}+6\frac{x^2}{y^2}+3\left(7-12\frac{x}{y}
 +5\frac{x^2}{y^2}-\frac{x^3}{y^3}+3\frac{x+y}{x-y}\right)\ln\frac{x}{y}\,\n\\
 &&+3\left(28-20\frac{x}{y}+7\frac{x^2}{y^2}-\frac{x^3}{y^3}\right)
 \frac{f(x,x^2-4xy)}{x-4y}\,,
 \label{eq:PV}
\eea
with the expression of the function $f$ as, 
\bea
 f(a,b)=
 \begin{cases}
   \sqrt{b}\ln{\left|\frac{a-\sqrt{b}}{a+\sqrt{b}}\right|}\,
   \hspace{1.7cm}\rm{for}\,\,b>0\,,\n\\
   0\,\hspace{3.56cm}\rm{for}\,\,b=0\,,\n\\
   2\sqrt{-b}\, \arctan\,\frac{\sqrt{-b}}{a}\,\qquad\rm{for}\,\,b<0\,.
 \end{cases}
\label{eq:f}
\eea
These functions stated in the Eqn. \ref{eq:PV} are widely used in precision electroweak analyses as the results from the Passarino-Veltman reduction of the loop integrals. 
The new scalar contributions beyond the SM give rise one-loop vacuum polarization diagrams. 
$F(x,y)$ emanates from the derivative of the scalar two-point function, $i.e.,$ the differences in the self-energies. 
This function encapsulates the outcome of the separation in the masses between two particles running in the loop.  The $\mathbb{T}$ parameter calculation employs this function. 
$G(x,y,z)$ and $\widehat{G}(x,y)$ exude from the momentum expansion of the vacuum polarization diagrams associating the scalars. 
These functions encase the issue of the new scalar masses on the transverse parts of the gauge boson self-energies. 
The $\mathbb{S,\,U,\,{\rm and}\,X}$ parameter calculations use the function $G(x,y,z)$, whereas $\mathbb{S}$ and $\mathbb{U}$ parameter calculations also utilize the function $\widehat{G}(x,y)$. 
$H(x,y,z)$ and $\widehat{H}(x,y)$ emerge from the second derivatives of the electroweak gauge bosons self-energies with respect to the momentum-squared at the one-loop level. 
These functions enclose the consequence of the curvature of the vacuum polarization functions and the modifications of the gauge-boson propagators due to the presence of the new scalars. 
The $\mathbb{V}$ and $\mathbb{W}$ parameter calculations engage the functions $H(x,y,z)$ and $\widehat{H}(x,y)$. 
For detailed derivation of these functions, one may consult \cite{Grimus:2008nb}. 

To express the oblique parameters for different BSM models concerning the rotation matrices has a key benefit as the knowledge of the components of the corresponding rotation matrices is enough to perceive the expression of the oblique parameters.
In this paper, to give the expressions of the
oblique parameters, we consider the scalar sectors of the models, such that, all the neutral scalars have VEVs, and all of them mix with each other. 
However, in the distinct cases, with some neutral scalars lacking VEVs, these oblique parameters formulations are valid as long as the rotation matrices for the scalar sector of the corresponding BSM model are changed accordingly.

We acquire the suitable method to formulate the oblique parameter $\mathbb{O}$ as $\bar{\mathbb{O}}$, where we multiply separate oblique parameters with separate prefactors to part with,
\bea
 \bar{\mathbb{S}}\equiv\frac{\a}{4s_W^2c_W^2}\mathbb{S}\,,\quad\quad
 \bar{\mathbb{T}}\equiv\a \mathbb{T}\,,\quad\quad
 \bar{\mathbb{U}}\equiv\frac{\a}{4s_W^2}\mathbb{U}\,,\quad\quad
 \bar{\mathbb{V}}\equiv\a \mathbb{V}\,,\quad\quad
 \bar{\mathbb{W}}\equiv\a \mathbb{W}\,,\quad\quad
 \bar{\mathbb{X}}\equiv\frac{\a}{s_Wc_W}\mathbb{X}\,,
 \label{eq:rescale}
\eea
with, $\theta_W$ as weak mixing angle, and $s_W$, $c_W$ as sine and cosine of it respectively.
Note that, the rescaling of the oblique parameters in the Eqn. [\ref{eq:rescale}] introduces no new information, though it is a common practice in some papers \cite{Grimus:2007if, Grimus:2008nb} to state the oblique parameters in this way. 

Ensuing \cite{Peskin:1991sw, 0802.4353ref1, 0802.4353ref2,
0802.4353ref3}, we sum up the oblique parameters :

\bea
 &&\bar{\mathbb{S}}=\frac{A_{ZZ}(m_Z^2)-A_{ZZ}(0)}{m_Z^2}
 - \frac{\p A_{\gamma\gamma}(q^2)}{\p q^2}\bigg|_{q^2=0} +\,\,
 \frac{c_W^2-s_W^2}{c_W s_W}\frac{\p A_{\gamma Z}(q^2)}{\p q^2}
 \bigg|_{q^2=0}\,,\n\\
 &&\bar{\mathbb{T}}=\frac{A_{WW}(0)}{m_W^2}-\frac{A_{ZZ}(0)}{m_Z^2}\,,\n\\
 &&\bar{\mathbb{U}}=\frac{A_{WW}(m_W^2)-A_{WW}(0)}{m_W^2}
 -c_W^2\frac{A_{ZZ}(m_Z^2)-A_{ZZ}(0)}{m_Z^2}\,
 -s_W^2\frac{\p A_{\gamma\gamma}(q^2)}{\p q^2}\bigg|_{q^2=0}
 +2c_Ws_W\frac{\p A_{\gamma Z}(q^2)}{\p q^2}\bigg|_{q^2=0}\,,\n\\
 &&\bar{\mathbb{V}}=\frac{\p A_{ZZ}(q^2)}{\p q^2}\bigg|_{q^2=m_Z^2}
 -\frac{A_{ZZ}(m_Z^2)-A_{ZZ}(0)}{m_Z^2}\,,\n\\
 &&\bar{\mathbb{W}}=\frac{\p A_{WW}(q^2)}{\p q^2}\bigg|_{q^2=m_W^2}
 -\frac{A_{WW}(m_W^2)-A_{WW}(0)}{m_W^2}\,,\n\\
 &&\bar{\mathbb{X}}=\frac{\p A_{\gamma Z}(q^2)}{\p q^2}\bigg|_{q^2=0}
 -\frac{A_{\gamma Z}(m_Z^2)}{m_Z^2}\,,\hspace{11cm}
 \label{eq:Obar}
\eea
with, $A_{VV^{'}}(q^2)$ being the coefficient of $g^{\mu\nu}$ in the vacuum polarization tensor $\Pi^{\mu\nu}_{VV^{'}}(q)$, which is a
function of the four momentum $(q^{\a})$ of the gauge bosons, given by,
$
 \Pi^{\mu\nu}_{VV^{'}}(q)=g^{\mu\nu}A_{VV^{'}}(q^2) 
 + q^{\mu}q^{\nu}B_{VV^{'}}(q^2)\,.
$
$VV^{'}$ can be anything of $ZZ$, $\gamma\gamma$, $\gamma Z$, and $WW$.

In all the three BSM models (Rx2SM in the section \ref{Rx2SM model}, N2HDM in the section \ref{N2HDM model}, and 3HDM in the section \ref{3HDM model}) with extended scalar sectors considered in this paper, these tensors collect one-loop contributions from the physical scalars through their couplings to $VV^{\prime}$. 
The kinetic parts of the Lagrangian of these three models (Eqns.\,\ref{eq:Lkin_Rx2SM}, \ref{eq:Lkin_N2HDM}, \ref{eq:Lkin_3HDM}) accord to the ratio of the couplings between the CP-even neutral Higgs bosons $H_{1,\,2,\,3}$ in the BSM model and the gauge bosons $W,\,Z$ to that in the SM (Eqns. \ref{eq:HVV_Rx2SM}, \ref{eq:HVV_N2HDM}, \ref{eq:HVV_3HDM}), whereas the Yukawa parts of the Lagrangians of these three BSM models (Eqns. \ref{eq:Yukawa_Rx2SM}, \ref{eq:Yukawa_N2HDM}, \ref{eq:Yukawa_3HDM}) commit to the ratio of the couplings between the CP-even neutral Higgs bosons in the BSM models and the fermionic pairs to that in the SM (Eqn. \ref{eq:Hff_Rx2SM}, Tab. \ref{tab:N2HDM}, Tab. \ref{tab:3HDM}). 
The specifications of these scale factors of the couplings for each model completely establish the connection between the CP-even neutral scalars of the specified three BSM models and the vacuum polarization tensors, without the requirements of the explicit calculations of the loop integrals.  

Now, let us quote the expressions of the oblique parameters $\overline{\mathbb{O}}$ given in the Eqn. \ref{eq:Obar} from the reference \cite{Grimus:2008nb} in terms of the matrices $\cal{U},\, \cal{V}$, the mass $m_a$ of the charged scalars $S_a^{\pm}$, the mass $\mu_b$ of the neutral scalars $S_b^0$, the functions $F,\,G,\,\widehat{G},\,H,\,\widehat{H}$ as listed in the Eqn. \ref{eq:f} : 
\bea
 \bar{\mathbb{S}} &=& 
 \frac{g^2}{384\pi^2c_W^2} \Big\{ \sum_{a=2}^{n_D} [2s_W^2-({\cal{U}}^{\dagger}{\cal{U}})_{aa}]^2 G(m_a^2,m_a^2,m_Z^2) 
 + 
 2 \sum_{a=2}^{n_D-1} \sum_{a^{\prime}=a+1}^{n_D} |{\cal{U}}^{\dagger}{\cal{U}})_{aa^{\prime}}|^2 G(m_a^2,m_{a^{\prime}}^2,m_Z^2) \nonumber\\
 &+ &
 \sum_{b=2}^{m-1} \sum_{b^{\prime}=b+1}^m [ {\rm Im} ({\cal{V}}^{\dagger}{\cal{V}})_{bb^{\prime}} ]^2 G(\mu_b^2,\mu_{b^{\prime}}^2,m_Z^2) 
 - 2 \sum_{a=2}^{n_D} ({\cal{U}}^{\dagger}{\cal{U}})_{aa} \,{\rm ln}\, m_a^2 
 + \sum_{b=2}^m ({\cal{V}}^{\dagger}{\cal{V}})_{bb} \,{\rm ln}\, \mu_b^2 
 - \,{\rm ln}\, m_H^2 \nonumber\\
 &+&
 \sum_{b=2}^m[ {\rm Im} ({\cal{V}}^{\dagger}{\cal{V}})_{1b} ]^2 \widehat{G} (\mu_b^2,m_Z^2) 
 - \widehat{G} (m_H^2,m_Z^2) \Big\}
 \label{eq:Sbar}
\eea
\bea
 \overline{\mathbb{T}} &=& 
 \frac{g^2}{64\pi^2 m_W^2} 
 \Big\{ 
 \sum_{a=2}^{n_D} \sum_{b=2}^m |({\cal{U}}^{\dagger}{\cal{V}})_{ab}|^2 F(m_a^2,\mu_b^2) 
 - 
 \sum_{b=2}^{m-1} \sum_{b^{\prime}=b+1}^m [ {\rm Im} ({\cal{V}}^{\dagger}{\cal{V}})_{bb^{\prime}} ]^2 F(\mu_b^2,\mu_{b^{\prime}}^2) \nonumber\\
 &-& 
 2 \sum_{a=2}^{n_D-1} \sum_{a^{\prime}=a+1}^{n_D} |({\cal{U}}^{\dagger}{\cal{U}})_{aa^{\prime}}|^2 F(m_a^2,m_{a^{\prime}}^2) 
 + 3 \sum_{b=2}^m [ {\rm Im} ({\cal{V}}^{\dagger}{\cal{V}})_{1b} ]^2 [ F(m_Z^2,\mu_b^2) - F(m_W^2,\mu_b^2) ] \nonumber\\
 &-& 
 3 [ F(m_Z^2,m_H^2) - F(m_W^2,m_H^2) ]
 \Big\}
 \label{eq:Tbar}
\eea
\bea
 \overline{\mathbb{U}} &=& 
 \frac{g^2}{384\pi^2} 
 \Big\{
 \sum_{a=2}^{n_D} \sum_{b=2}^m |({\cal{U}}^{\dagger}{\cal{V}})_{ab}|^2 G(m_a^2,\mu_b^2,m_W^2) 
 - \sum_{a=2}^{n_D} [ 2s_W^2 -({\cal{U}}^{\dagger}{\cal{U}})_{aa} ]^2 G(m_a^2,m_a^2,m_Z^2) \nonumber\\
 &-&
 2 \sum_{a=2}^{n_D-1} \sum_{a^{\prime}=a+1}^{n_D} |({\cal{U}}^{\dagger}{\cal{U}})_{aa^{\prime}}|^2 G(m_a^2,m_{a^{\prime}}^2,m_Z^2) 
 - \sum_{b=2}^{m-1} \sum_{b^{\prime}=b+1}^{m} [ {\rm Im} ({\cal{V}}^{\dagger}{\cal{V}})_{bb^{\prime}} ]^2 G(\mu_b^2,\mu_{b^{\prime}}^2,m_Z^2) \nonumber\\
 &+& 
 \sum_{b=2}^m [ {\rm Im} ({\cal{V}}^{\dagger}{\cal{V}})_{1b} ]^2 \big[ \widehat{G}(\mu_b^2,m_W^2) - \widehat{G}(\mu_b^2,m_Z^2) \big] 
 - \widehat{G}(m_H^2,m_W^2) 
 + \widehat{G}(m_H^2,m_Z^2) 
 \Big\} 
 \label{eq:Ubar}
\eea
\bea
 \overline{\mathbb{V}} &=&
 \frac{g^2}{384\pi^2 c_W^2} 
 \Big\{
 \sum_{a=2}^{n_D} [ 2s_W^2 - ({\cal{U}}^{\dagger}{\cal{U}})_{aa} ]^2 H(m_a^2,m_a^2,m_Z^2) 
 + 
 2 \sum_{a=2}^{n_D-1} \sum_{a^{\prime}=a+1}^{n_D} |({\cal{U}}^{\dagger}{\cal{U}})_{aa^{\prime}}|^2 H(m_a^2,m_{a^{\prime}}^2,m_Z^2) \nonumber\\
 &+& 
 \sum_{b=2}^{m-1} \sum_{b^{\prime}=b+1}^m [ {\rm Im} ({\cal{V}}^{\dagger}{\cal{V}})_{bb^{\prime}} ]^2 H(\mu_b^2,\mu_{b^{\prime}}^2,m_Z^2) 
 + 
 \sum_{b=2}^m [ {\rm Im} ({\cal{V}}^{\dagger}{\cal{V}})_{1b} ]^2 \widehat{H} (\mu_b^2,m_Z^2) 
 - 
 \widehat{H}(m_H^2,m_Z^2) 
 \Big\}
 \label{eq:Vbar}
\eea
\bea
 \overline{\mathbb{W}} &=& 
 \frac{g^2}{384\pi^2} 
 \Big\{ 
 \sum_{a=2}^{n_D} \sum_{b=2}^m |({\cal{U}}^{\dagger}{\cal{V}})_{ab}|^2 H(m_a^2,\mu_b^2,m_W^2) 
 + 
 \sum_{b=2}^m [ {\rm Im} ({\cal{V}}^{\dagger}{\cal{V}})_{1b} ]^2 \widehat{H} (\mu_b^2,m_W^2) 
 - 
 \widehat{H} (m_H^2,m_W^2) 
 \Big\} 
 \label{eq:Wbar}
\eea
\bea
 \overline{\mathbb{X}} &=&
 -\frac{g^2 s_W}{192 \pi^2 c_W} 
 \sum_{a=2}^{n_D} [ 2s_W^2 - ({\cal{U}}^{\dagger}{\cal{U}})_{aa} ] G(m_a^2,m_a^2,m_Z^2)
 \label{eq:Xbar}
\eea
We call up the formulations of oblique parameters for the three models in the following three subsections.
For the models, where all the neutral scalars mix with each other, all the components of the mixing matrix are non-zero.
Considering the models, with one of the BSM neutral scalars lacking VEV, the resembling matrix gets reduced, as four of its components vanish, and the other becomes unity.

To probe the parameter spaces of the specified BSM models of our interest, the experimentally available values of the oblique parameters \cite{Coleppa:2025qst} are also required, and we list here the first three oblique parameters :
\be
 \mathbb{S} = -0.02\,\pm\,0.10\,,
 \quad
 \mathbb{T} = 0.03\,\pm\,0.12\,,
 \quad
 \mathbb{U} = 0.01\,\pm\,0.11\,.
\label{eq:STU_experiment}
\ee
Though the previous equations (\ref{eq:Obar} -- \ref{eq:Xbar}) express the oblique parameters in a rescaled way, as given in the Eqn. \ref{eq:rescale}, the results for the three BSM models in the next subsections provide the expressions without the rescaled manner.

\subsection{Rx2SM}
\label{Rx2SM oblique}

\numberwithin{equation}{subsection}

In this subsection, we enlist the oblique parameters for the Two Real 
Singlet scalars extended SM ($Rx2SM$), in terms of the rotation matrix 
$O_\a$. 

\bea
 \mathbb{S} &=& \frac{4s_W^2c_W^2}{\a} \frac{g^2}{384\pi^2c_W^2}\left[\sum_{j=1}^3 O_{{\a}_{j1}}^2 
 \left(\ln m_{H_j}^2+\hat{G}(m_{H_j}^2,m_Z^2)\right) 
 - \left(\ln m_{H_{SM}}^2+\hat{G}(m_{H_{SM}}^2,m_Z^2)\right) \right]\,,\n\\
 \mathbb{T} &=& \frac{1}{\a} \frac{3g^2}{64\pi^2m_W^2}\Bigg[ \sum_{j=1}^3
 O_{\a_{j1}}^2 \left(F(m_Z^2,m_{H_j}^2)-F(m_W^2,m_{H_j}^2)\right)
 - \left(F(m_Z^2,m_{H_{SM}}^2)-F(m_W^2,m_{H_{SM}}^2)\right)
 \Bigg]\,,\n\\
 \mathbb{U} &=& \frac{4s_W^2}{\a} \frac{g^2}{384\pi^2}\Bigg[ \sum_{j=1}^3 O_{\a_{j1}}^2
 \left(\hat{G}\left(m_{H_j}^2,m_W^2\right)-\hat{G}\left(m_{H_j}^2,m_Z^2\right)
 \right)- \left(\hat{G}\left(m_{H_{SM}}^2,m_W^2\right)-\hat{G}
 \left(m_{H_{SM}}^2,m_Z^2\right)\right)\Bigg]\,,\n\\
 \mathbb{V} &=& \frac{1}{\a} \frac{g^2}{384\pi^2c_W^2}\Bigg[ \sum_{j=1}^3
 O_{\a_{j1}}^2 \hat{H}\left(m_{H_j}^2,m_Z^2\right)
 - \hat{H}\left(m_{H_{SM}}^2,m_Z^2\right) \Bigg]\,,\n\\
 \mathbb{W} &=& \frac{1}{\a} \frac{g^2}{384\pi^2}\Bigg[ \sum_{j=1}^3
 O_{\a_{j1}}^2 \hat{H}\left(m_{H_j}^2,m_W^2\right)
 - \hat{H}\left(m_{H_{SM}}^2,m_W^2\right) \Bigg]\,.
\eea
The above expressions of the oblique parameters satisfy the main goal of this paper. 
For completeness, we probe the allowed parameter spaces considering some of the oblique parameters within the range of $1\sigma$ or $2\sigma$ of the observed data. 
For this purpose, this paper only concentrates on the main three oblique parameters, $i.e.$, $\mathbb{S}$, $\mathbb{T}$, and $\mathbb{U}$. 
Any of the masses of the Higgs bosons $m_{H_j}$ can be equal to the mass of the SM Higgs boson $m_{H_{SM}}$. 
Here, we set $m_{H_2}=m_{H_{SM}}$, such that, $m_{H_1} < m_{H_{SM}} < m_{H_3}$. 
We scan over $m_{H_1}$ and $m_{H_3}$ between $63-120$ and $130 - 1000$ GeV, respectively. 
All the mixing angles $\a_{1,2,3}$ between the CP-even neutral scalars are scanned over the full region of $-\pi/2$ to $\pi/2$. 
Let us first consider the case where any of $\mathbb{S},\,\mathbb{T},\,{\rm and}\,\mathbb{U}$ be within the $1\sigma$ or $2\sigma$ region of the observed oblique parameter value. 
Fig.\,\ref{fig:Rx2SM_S_m1} shows such case for $1\sigma$ (left plot) as well as $2\sigma$ (right plot), where Y-axis indicate the value of the $\mathbb{S}$ parameter and X-axis indicate the mass $m_{H_1}$ of the lightest scalar in GeV, as an example. 
The black solid lines indicate the allowed region as per the observed data of the oblique parameter $\mathbb{S}$. 
The cyan, black, and violet points represent when only $\mathbb{S}$, $\mathbb{T}$, and $\mathbb{U}$ are within allowed region respectively. 
From this Fig.\,, one can clearly notice that the constraint on any of the oblique parameters can not assure the values of the $\mathbb{S}$ parameter within the limits. 
Similarly, it can be inferred that, the constrained values of any of these parameters do not always imply that the other oblique parameters will be surely within the range. 
Hence we consider all of the three oblique parameters are within the observed limit. 
We follow this for this Rx2SM model as well as for the other two models, $viz.$ N2HDM and 3HDM. 
Next, in Fig.\,\ref{fig:STU_m1m3}, all of the $\mathbb{S},\, \mathbb{T},\, and \, \mathbb{U}$ parameters are considered to be within the constrained region, either in $1\sigma$ (left plot) or in $2\sigma$ (right plot). 
The orange points (the boxes) represent the allowed parameter space in the $m_{H_3} - m_{H_1}$ plane of the Rx2SM model. 
The Fig.\,\ref{fig:STU_angle} shows the allowed parameter space in $\a_1 - \a_2$ (left plot) and $\a_1 - \a_3$ (right plot) plane. 
\subsection{N2HDM}
\label{N2HDM oblique}
%
\begin{figure}
 \begin{center}
 \includegraphics[width= 8.2cm]{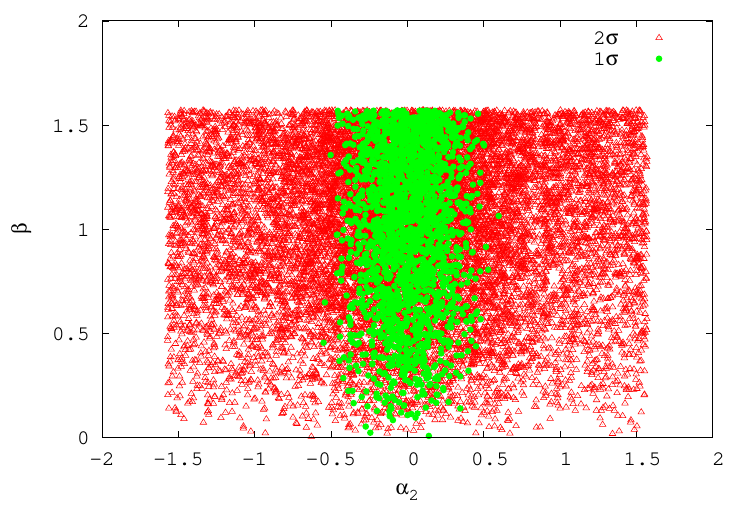} \ \
 \includegraphics[width= 8.2cm]{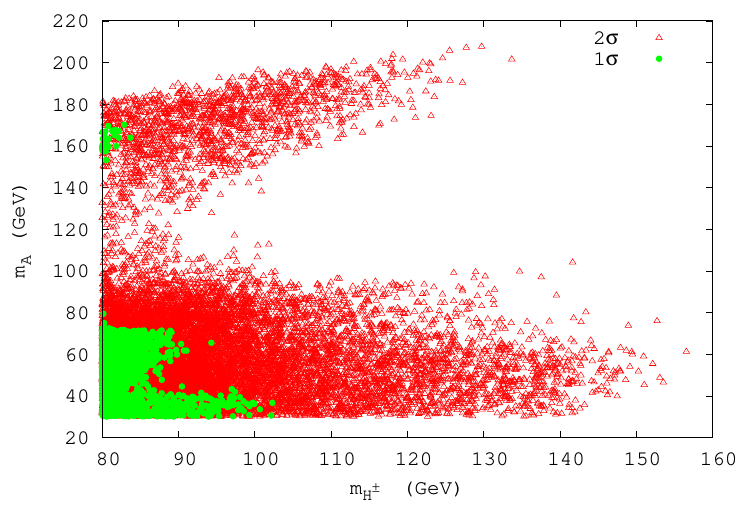}
 \end{center}
 \caption{\small Left plot : The mixing angle $\b$ as the function of the angle $\a_2$ in the model N2HDM.\\
 Right plot : The mass of the CP-odd neutral Higgs boson ($A$) as the function of the mass of charged scalar ($H^{\pm}$) in the model N2HDM.\\ 
 The triangles (red points), and the circles (green points) indicate the allowed region when all of the $\mathbb{S}$, $\mathbb{T}$, $\mathbb{U}$ parmeters are within $2\sigma$ and $1\sigma$, respectively. }
 \label{fig:STU_N2HDM}
 \end{figure}
Here, we procure the oblique parameters for N2HDM, concerning the rotation matrices
$O_\a$ and $O_\b$. 

\bea
 \hspace{-3cm}
 \mathbb{S} &=& \frac{4s_W^2c_W^2}{\a} \frac{g^2}{384\pi^2c_W^2}\Bigg[ \left(2s_W^2-1\right)^2 
 G\left(m_{H^+}^2,m_{H^+}^2,m_Z^2\right) + \left(\sum_{i=1}^2 O_{\b_{2i}}
 O_{\a_{1i}} \right)^2G\left(m_{H_1}^2,m_A^2,m_Z^2\right)\,\n\\
 \hspace{-3cm}
 &&+ \sum_{r=2}^3\left(\sum_{i=1}^2 O_{\b_{2i}} O_{\a_{ri}}\right)^2 
 G\left(m_A^2,m_{H_r}^2,m_Z^2\right) -2\ln m_{H^+}^2 +\sum_{j=1}^3\left(
 \sum_{i=1}^2 O_{\a_{ji}}^2\right)\ln m_{H_j}^2 
 + \ln m_A^2 \,\n\\
 \hspace{-3cm}
 &&+ \sum_{j=1}^3\left(\sum_{i=1}^2 O_{\b_{1i}}O_{\a_{ji}} 
 \right)^2\hat{G}\left(m_{H_j}^2,m_Z^2\right) 
 - \left(\ln m_{H_{SM}}^2\,+ \hat{G}\left(m_{H_{SM}}^2,m_Z^2\right)\right)
 \Bigg]\,,\n
\eea

\bea
\hspace{-1cm}
 \mathbb{T} &=& \frac{1}{\a} \frac{g^2}{64\pi^2m_W^2}\Bigg[ \sum_{j=1}^3 \bigg| 
 \sum_{i=1}^2 O_{\b_{2i}} O_{\a_{ji}} \bigg|^2 F\left(m_{H^+}^2,
 m_{H_j}^2\right) + F\left(m_{H^+}^2,m_A^2\right)
 -\left(\sum_{i=1}^2 O_{\b_{2i}} O_{\a_{1i}}\right)^2 F\left( 
 m_{H_1}^2,m_A^2\right) \n\\
 \hspace{-1cm}&&
 - \sum_{r=2}^3\left(\sum_{i=1}^2 O_{\b_{2i}} 
 O_{\a_{ri}}\right)^2 F\left(m_A^2,m_{H_r}^2\right)
 +3\sum_{j=1}^3\left(\sum_{i=1}^2 O_{\b_{1i}} O_{\a_{ji}}\right)^2 
 \left(F\left(m_Z^2,m_{H_j}^2\right)-F\left(m_W^2,m_{H_j}^2\right)\right)
 \n\\ \hspace{-1cm}
 &&- 3 \left(F\left(m_Z^2,m_{H_{SM}}^2\right) 
 - F\left(m_W^2,m_{H_{SM}}^2\right)\right)\Bigg]\,,\n
\eea

\bea
 \mathbb{U} &=& \frac{4s_W^2}{\a} \frac{g^2}{384\pi^2}\Bigg[ \sum_{j=1}^3\left|\sum_{i=1}^2 
 O_{\b_{2i}} O_{\a_{ji}}\right|^2 G\left(m_{H^+}^2,m_{H_j}^2,m_W^2\right) 
 + G\left(m_{H^+}^2,m_A^2,m_W^2\right)- \left(2s_W^2-1\right)^2
 \n\\ \hspace{-1cm}
 && G\left(m_{H^+}^2,m_{H^+}^2,m_Z^2\right) - \left(\sum_{i=1}^2 O_{\b_{2i}} O_{\a_{1i}}\right)^2 G\left(
 m_{H_1}^2,m_A^2,m_Z^2\right) 
 - \sum_{r=2}^3 \left(\sum_{i=1}^2 O_{\b_{2i}} O_{\a_{ri}}\right)^2
 G\left(m_A^2,m_{H_r}^2,m_Z^2\right)\n\\ \hspace{-1cm} 
 &&+ \sum_{j=1}^3 \left(\sum_{i=1}^2 
 O_{\b_{1i}} O_{\a_{ji}}\right)^2 \left(\hat{G}\left(m_{H_j}^2,m_W^2\right)
 - \hat{G}\left(m_{H_j}^2,m_Z^2\right)\right)
 - \left(\hat{G}\left(m_{H_{SM}}^2,m_W^2\right) 
 - \hat{G}\left(m_{H_{SM}}^2,m_Z^2\right)\right) \Bigg]\,,\n
\eea

\bea
 \mathbb{V} &=& \frac{1}{\a} \frac{g^2}{384\pi^2c_W^2}\Bigg[ \left(2s_W^2-1\right)^2
 H\left(m_{H^+}^2,m_{H^+}^2,m_Z^2\right) + \left(\sum_{i=1}^2 
 O_{\b_{2i}} O_{\a_{1i}}\right)^2 H\left(m_{H_1}^2,m_A^2,m_Z^2\right)\n\\ \hspace{-1cm}
 && +\sum_{r=2}^3 \left(\sum_{i=1}^2 O_{\b_{2i}} O_{\a_{ri}}\right)^2
 H\left(m_A^2,m_{H_r}^2,m_Z^2\right) + \sum_{j=1}^3\left(
 \sum_{i=1}^2 O_{\b_{1i}} O_{\a_{ji}}\right)^2 \hat{H}\left(m_{H_j}^2,
 m_Z^2\right)
 -\hat{H}\left(m_{H_{SM}}^2,m_Z^2\right)\Bigg]\,,\n
\eea

\bea
 \mathbb{W} &=& \frac{1}{\a} \frac{g^2}{384\pi^2}\Bigg[ \sum_{j=1}^3 \left(\sum_{i=1}^2
 O_{\b_{2i}} O_{\a_{ji}}\right)^2 H\left(m_{H^+}^2,m_{H_j}^2,m_W^2\right) 
 + H\left(m_{H^+}^2,m_A^2,m_W^2\right)\n\\\hspace{-1cm} 
 && + \sum_{j=1}^3 \left(\sum_{i=1}^2 O_{\b_{1i}} O_{\a_{ji}}\right)^2
 \hat{H}\left(m_{H_j}^2,m_W^2\right) - \hat{H}\left(m_{H_{SM}}^2,m_W^2
 \right) \Bigg]\,,\n
\eea

\bea
 \mathbb{X} &=& - \frac{s_Wc_W}{\a} \frac{g^2s_W}{192\pi^2c_W}\Bigg[ \left(2s_W^2-1\right)
 G\left(m_{H^+}^2,m_{H^+}^2,m_Z^2\right) \Bigg]\,.
\eea
Here, we set $m_{H_2}=m_{H_{SM}}$, such that, $m_{H_1} < m_{H_{SM}} < m_{H_3}$. 
We scan over $m_{H_1}$ between $63-120$ GeV, $m_{H_3}$ between $130 - 1000$ GeV, $m_{H^{\pm}}$ between $80 - 200$ GeV, and $m_A$ between $30 - 250$ GeV. 
All the mixing angles $\a_{1,2,3},\,\b$ are scanned over the full region of $-\pi/2$ to $\pi/2$. 
In the Fig.\,\ref{fig:STU_m1m3}, all of the $\mathbb{S},\, \mathbb{T},\, {\rm and} \, \mathbb{U}$ parameters are considered to be within the constrained region, either in $1\sigma$ (left plot) or in $2\sigma$ (right plot). 
The pink points (the circles) represent the allowed parameter space in the $m_{H_3} - m_{H_1}$ plane of the N2HDM model. 
For $1\sigma$ region, $m_{H_1}$ and $m_{H_3}$ are below $85$ and $200$ GeV, whereas for $2\sigma$ region, there are no such limit can be seen from this Fig. 
Next, the Fig.\,\ref{fig:STU_angle} shows the allowed parameter space in $\a_1 - \a_2$ (left plot) and $\a_1 - \a_3$ (right plot) plane, when all of the $\mathbb{S},\, \mathbb{T},\, \mathbb{U}$ parameters are within $1\sigma$. 
From this, one can see that, there is no limit on the angle $\a_1$, whereas the value of $\a_2$ is limited within $-0.75\, {\rm to}\, 0.75$ and $\a_3$ is beyond this range. 
There is no such limit on these angles $\a_{1,2,3}$ for the oblique parameters within the range of $2\sigma$, and hence all of these are not shown in this paper. 
The left plot of the Fig.\,\ref{fig:STU_N2HDM} shows that there is no limit on $\a_2$ for $2\sigma$ $\mathbb{S},\, \mathbb{T},\, \mathbb{U}$. 
The Y-axis of the same plot shows that there is no limit on the range of $\b$ for these oblique parameters in $1\sigma$ (green circles) or $2\sigma$ (red triangles) regions. 
The right plot of this Fig. depict the allowed parameter space in the $m_A - m_{H^{\pm}}$ plane with the same color scheme. 
\subsection{3HDM}
\label{3HDM oblique}
\begin{figure}
 \begin{center}
  \includegraphics[width= 8.2cm]{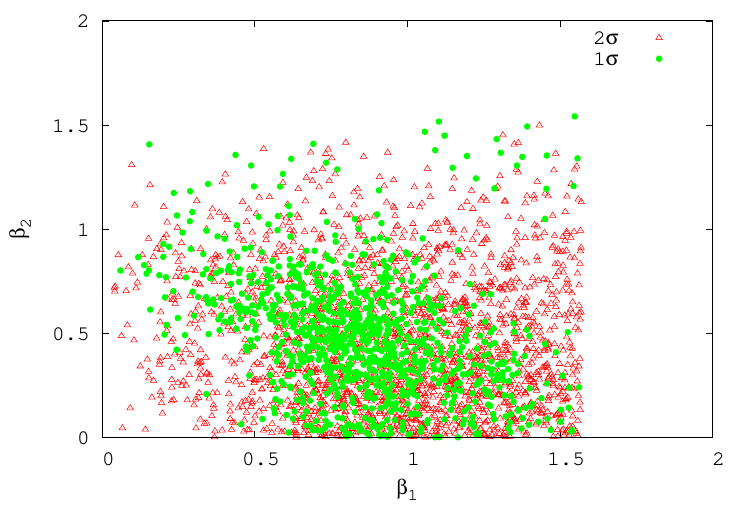} \ \
 \includegraphics[width= 8.2cm]{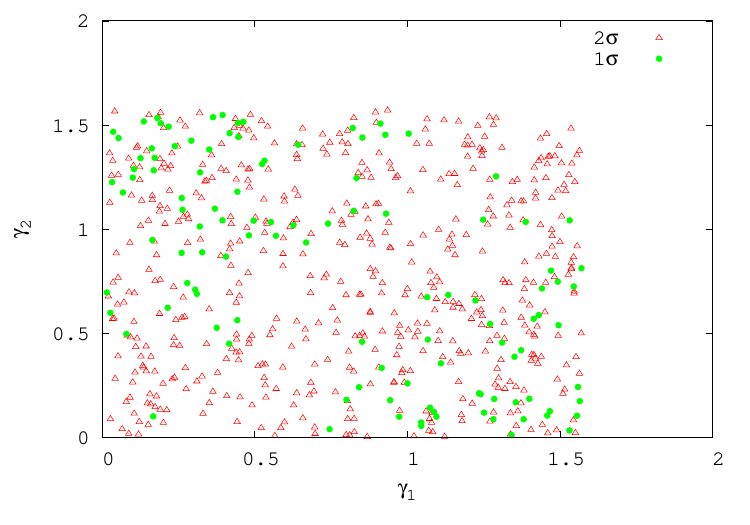}
 \end{center}
 \caption{\small Left plot : The $\b_1 - \b_2$ plane, Right plot : The $\gamma_1 - \gamma_2$ plane, in the model 3HDM.\\
 The triangles (red points), and the circles (green points) indicate the allowed region when all of the $\mathbb{S}$, $\mathbb{T}$, $\mathbb{U}$ parmeters are within $2\sigma$ and $1\sigma$, respectively. }
 \label{fig:STU_b1b2g1g2}
 \end{figure}
\begin{figure}
 \begin{center}
  \includegraphics[width= 8.2cm]{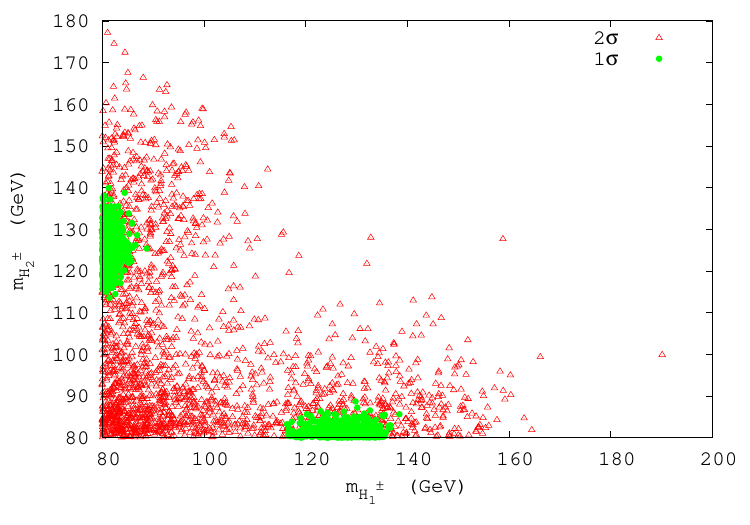} \ \
 \includegraphics[width= 8.2cm]{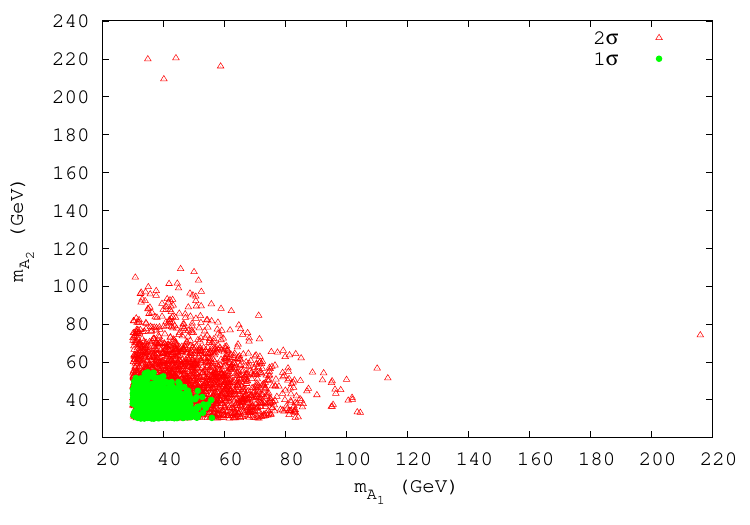}
 \end{center}
 \caption{\small The allowed masses of the CP-odd neutral Higgs bosons ($A_{1,2}$) and the charged scalars ($H_{1,2}^{\pm}$) in the model 3HDM, in the right and left plot, respectively. The triangles (red points), and the circles (green points) indicate the allowed region in the mass planes when all of the $\mathbb{S}$, $\mathbb{T}$, $\mathbb{U}$ parmeters are within $2\sigma$ and $1\sigma$, respectively. }
 \label{fig:STU_3HDM}
 \end{figure}
In this subsection, we enlist the oblique parameters for the Three Higgs Doublet Model ($3HDM$), concerning the rotation matrices
$O_\a$, $O_{\b\gamma_1}$, and $O_{\b\gamma_2}$. 

\bea
 \hspace{-1cm}
 \mathbb{S} &=& \frac{4s_W^2c_W^2}{\a} \frac{g^2}{384\pi^2c_W^2} \Bigg[ \left(2s_W^2-1\right)^2
 \sum_{i=1}^2 G\left(m_{H_i^+}^2,m_{H_i^+}^2,m_Z^2\right)\, 
  + \sum_{r=2}^3\sum_{i=1}^{r-1}\left(\sum_{j=1}^3 
 O_{{\b\gamma_1}_{rj}} O_{\a_{ij}}\right)^2 G\left(m_{H_i}^2,m_{A_{r-1}}^2,m_Z^2\right)
 \,\n\\ \hspace{-1cm}
 && + \sum_{r=2}^3\sum_{i=r}^3\left(\sum_{j=1}^3 
 O_{{\b\gamma_1}_{rj}} O_{\a_{ij}}\right)^2 
 G\left(m_{A_{r-1}}^2,m_{H_i}^2,m_Z^2\right)\,
 -2\sum_{i=1}^2\ln m_{H_i^+}^2 
 +\sum_{j=1}^3\left(\sum_{k=1}^3 O_{\a_{jk}}^2\right)\ln m_{H_j}^2 
 +\sum_{i=1}^2 \ln m_{A_i}^2 \n\\
 \hspace{-1cm}
 &&+ \sum_{j=1}^3\left(\sum_{k=1}^3 O_{{\b\gamma_1}_{1k}}O_{\a_{jk}}\right)^2 
 \hat{G}\left(m_{H_j}^2,m_Z^2\right) 
 - \left(\ln m_{H_{SM}}^2 + \hat{G}\left(m_{H_{SM}}^2,m_Z^2\right)\right)
 \Bigg]\,,\n
\eea

\bea
 \mathbb{T} &=& \frac{1}{\a} \frac{g^2}{64\pi^2m_W^2}\Bigg[
 \sum_{i=1}^2 \sum_{j=1}^3 \bigg| \sum_{k=1}^3 O_{{\b\gamma_2}_{k,i+1}} 
 O_{\a_{jk}} \bigg|^2 F\left(m_{H_i^+}^2,m_{H_j}^2\right)\,
 + \sum_{i=1}^2 \sum_{r=2}^3 \bigg|\sum_{k=1}^3 O_{{\b\gamma_1}_{rk}}
 O_{{\b\gamma_2}_{k,i+1}} \bigg|^2 
 F\left(m_{H_i^+}^2,m_{A_{r-1}}^2\right)\n\\ \hspace{-1cm}
 && -\sum_{r=2}^3\sum_{i=1}^{r-1}\left(\sum_{j=1}^3 O_{{\b\gamma_1}_{r,j}} 
 O_{\a_{ij}}\right)^2 F\left(m_{H_i}^2,m_{A_{r-1}}^2\right)
 - \sum_{r=2}^3\sum_{i=r}^3\left(\sum_{j=1}^3 O_{{\b\gamma_1}_{rj}} 
 O_{\a_{ij}}\right)^2 F\left(m_{A_{r-1}}^2,m_{H_i}^2\right)\n\\ \hspace{-1cm}
 && +3\sum_{j=1}^3\left(\sum_{k=1}^3 O_{{\b\gamma_1}_{1k}} 
 O_{\a_{jk}}\right)^2 \left(F\left(m_Z^2,m_{H_j}^2\right)
 -F\left(m_W^2,m_{H_j}^2\right)\right) 
 - 3 \left(F\left(m_Z^2,m_{H_{SM}}^2\right) 
 - F\left(m_W^2,m_{H_{SM}}^2\right)\right)\Bigg]\,,\n\\
 \hspace{-1cm}
 \mathbb{U} &=& \frac{4s_W^2}{\a} \frac{g^2}{384\pi^2}\Bigg[
 \sum_{i=1}^2 \sum_{j=1}^3 \bigg| \sum_{k=1}^3 O_{{\b\gamma_2}_{k,i+1}} 
 O_{\a_{jk}} \bigg|^2 G\left(m_{H_i^+}^2,m_{H_j}^2,m_W^2\right)
 + \sum_{i=1}^2 \sum_{r=2}^3 \bigg|\sum_{k=1}^3 O_{{\b\gamma_1}_{rk}}
 O_{{\b\gamma_2}_{k,i+1}} \bigg|^2 
 \n\\ \hspace{-1cm}
 && G\left(m_{H_i^+}^2,m_{A_{r-1}}^2,m_W^2\right) - \left(2s_W^2-1\right)^2 \sum_{i=1}^2 G\left(m_{H_i^+}^2,
 m_{H_i^+}^2,m_Z^2\right) 
 -\sum_{r=2}^3 \sum_{i=1}^{r-1} \left(\sum_{j=1}^3 
 O_{{\b\gamma_1}_{rj}} O_{\a_{ij}}\right)^2 
 \n\\ \hspace{-1cm}
 && G\left( m_{H_i}^2,m_{A_{r-1}}^2,m_Z^2\right) -\sum_{r=2}^3\sum_{i=r}^3 \left(\sum_{j=1}^3
 O_{{\b\gamma_1}_{rj}} O_{\a_{ij}} \right)^2 
 G\left( m_{A_{r-1}}^2,m_{H_i}^2,m_Z^2\right)
 + \sum_{j=1}^3 \left(\sum_{k=1}^3 
 O_{{\b\gamma_1}_{1k}} O_{\a_{jk}}\right)^2 
 \n\\ \hspace{-1cm}
 && \left(\hat{G}\left(m_{H_j}^2,m_W^2\right)
 - \hat{G}\left(m_{H_j}^2,m_Z^2\right)\right) - \left(\hat{G}\left(m_{H_{SM}}^2,m_W^2\right)
 - \hat{G}\left(m_{H_{SM}}^2,m_Z^2\right)\right) 
 \Bigg]\,,\n
\eea

\bea
 \mathbb{V} &=& \frac{1}{\a} \frac{g^2}{384\pi^2c_W^2}\Bigg[ \left(2s_W^2-1\right)^2
 \sum_{i=1}^2 H\left(m_{H_i^+}^2,m_{H_i^+}^2,m_Z^2\right)
 +\sum_{r=2}^3\sum_{i=1}^{r-1} \left(\sum_{j=1}^3 O_{{\b\gamma_1}_{rj}} 
 O_{\a_{ij}}\right)^2 H\left(m_{H_i}^2,m_{A_{r-1}}^2,m_Z^2\right)\n\\
 \hspace{-1cm} 
 && +\sum_{r=2}^3\sum_{i=r}^3 \left(\sum_{j=1}^3 
 O_{{\b\gamma_1}_{rj}} O_{\a_{ij}}\right)^2 
 H\left(m_{A_{r-1}}^2,m_{H_i}^2,m_Z^2\right)
 + \sum_{j=1}^3\left( \sum_{k=1}^3 
 O_{{\b\gamma_1}_{1k}} O_{\a_{jk}}\right)^2 
 \hat{H}\left(m_{H_j}^2,m_Z^2\right) 
 -\hat{H}\left(m_{H_{SM}}^2,m_Z^2\right)\Bigg]\,,\n
\eea

\bea
 \mathbb{W} &=& \frac{1}{\a} \frac{g^2}{384\pi^2}\Bigg[
 \sum_{i=1}^2 \sum_{j=1}^3 \bigg| \sum_{k=1}^3 O_{{\b\gamma_2}_{k,i+1}} 
 O_{\a_{jk}} \bigg|^2 H\left(m_{H_i^+}^2,m_{H_j}^2,m_W^2\right)
 + \sum_{i=1}^2 \sum_{r=2}^3 \bigg|\sum_{k=1}^3 O_{{\b\gamma_1}_{rk}}
 O_{{\b\gamma_2}_{k,i+1}} \bigg|^2\,
 \n\\ \hspace{-1cm}
 &&H\left(m_{H_i^+}^2,m_{A_{r-1}}^2,m_W^2\right)
 \sum_{j=1}^3 \left(\sum_{k=1}^3 O_{{\b\gamma_1}_{1k}} O_{\a_{jk}}\right)^2
 \hat{H}\left(m_{H_j}^2,m_W^2\right) - \hat{H}\left(m_{H_{SM}}^2,m_W^2
 \right) \Bigg]\,,\n\\
 \mathbb{X} &=& - \frac{s_Wc_W}{\a} \frac{g^2s_W}{192\pi^2c_W}\Bigg[ \left(2s_W^2-1\right)
 \sum_{i=1}^2 G\left(m_{H_i^+}^2,m_{H_i^+}^2,m_Z^2\right)
 \Bigg]\,.
\eea
Here, we set $m_{H_2}=m_{H_{SM}}$, such that, $m_{H_1} < m_{H_{SM}} < m_{H_3}$. 
We scan over $m_{H_1}$ between $63-120$ GeV, $m_{H_3}$ between $130 - 1000$ GeV, $m_{H_{1,2}^{\pm}}$ between $80 - 200$ GeV, and $m_{A_{1,2}}$ between $30 - 250$ GeV. 
All the mixing angles $\a_{1,2,3},\,\b,\, \gamma_{1,2}$ are scanned over the full region of $-\pi/2$ to $\pi/2$. 
In the Fig.\,\ref{fig:STU_m1m3}, all of the $\mathbb{S},\, \mathbb{T},\, {\rm and} \, \mathbb{U}$ parameters are considered to be within the constrained region, either in $1\sigma$ (left plot) or in $2\sigma$ (right plot). 
The blue points (the triangles) represent the allowed parameter space in the $m_{H_3} - m_{H_1}$ plane of the 3HDM model. 
For $1\sigma$ region, $m_{H_1}$ and $m_{H_3}$ are below $70$ and $140$ GeV, whereas for $2\sigma$ region, there are no such limit can be seen from this Fig. for $m_{H_1}$, but for $m_{H_3}$ it is limited below $210$ GeV. 
Next, the Fig.\,\ref{fig:STU_angle} shows the allowed parameter space in $\a_1 - \a_2$ (left plot) and $\a_1 - \a_3$ (right plot) plane, when all of the $\mathbb{S},\, \mathbb{T},\, \mathbb{U}$ parameters are within $1\sigma$. 
There is no such limit on these angles $\a_{1,2,3}$ for the oblique parameters within the range of $2\sigma$, and hence these are not shown in this paper. 
The left and right plots of the Fig.\,\ref{fig:STU_b1b2g1g2} show that the values of the $\b_{1,2}$ and the $\gamma_{1,2}$ are, respectively, within $0 - \pi/2$ and there is no limit on these angles for $1\sigma$ (green circles) or $2\sigma$ (red triangles) $\mathbb{S},\, \mathbb{T},\, \mathbb{U}$. 
The Fig.\,\ref{fig:STU_3HDM} depict the allowed region in the $m_{H_1^{\pm}} - m_{H_2^{\pm}}$ (left plot) and $m_{A_1} - m_{A_2}$ (right plot) planes in the 3HDM with the same color scheme. 
This Fig. shows that, for $1\sigma$, $m_{A_{1,2}}$ are below $60$ GeV, mass of one of the $H_{1,2}^{\pm}$ is within $115 - 140$ GeV with the other charged Higgs mass within $80 - 90$ GeV, and for $2\sigma$, $m_{A_{1,2}}$ are below $220$ GeV, $m_{H_1}^{\pm}$ is below $200$ GeV, $m_{H_2}^{\pm}$ is below $180$ GeV. 
\section{Conclusions}
\label{Conclusions}

Considering three BSM scripts, each having three CP-even neutral scalars, we cipher $\mathbb{S}$, $\mathbb{T}$, $\mathbb{U}$, $\mathbb{V}$, $\mathbb{W}$, and $\mathbb{X}$. 
We consider the extension of the SM with two singlet scalars that are real, with one complex fellow and one real singlet scalar, and eventually with two complex scalars. 
We give the expressions, considering all the CP-even neutral scalars having VEVs. 
As we've formerly bandied, we present the expressions of the oblique parameters in such a way that the knowledge of the mixing matrices for each model is enough to prize the full expression of the oblique parameters of that particular model. 
These expressions are valid for the cases, where all the CP- indeed neutral scalars have VEVs and blend with each other, as well as for special cases when one of the three CP- indeed neutral scalars doesn't retain VEV and the mixing between them is confined.

The Rx2SM, having no BSM- charged scalars, doesn't contribute to the oblique parameter $\mathbb{X}$. 
One thing is to be noted that, the expressions of the oblique parameters are relatively simpler for the extension of the SM with singlets than that with doublets, as anticipated. 
Among the three BSM models, we accounted, one can conclude that, the expressions are simplest for Two Real Singlet scalars extended SM, a little complicated for N2HDM, and toughest for 3HDM, which is also reflected in the plots.

\vspace{0.5cm}
 {\em {\bf Acknowledgements}} --- The author thanks Prof. Anirban Kundu 
 for useful discussions. The major part of this work is done independently, 
 without any financial support, as the author is an honorary fellow under the SERB grant CRG/2018/004889. The last part of this work is done with the financial support of the ANRF-NPDF scholarship with grant no. PDF/2022/001784.

\appendix\markboth{Appendix}{Appendix}
\renewcommand{\thesection}{\Alph{section}}
\numberwithin{equation}{section}
\section{Mixing Matrices}
\label{A}

Below we list down the most general case in all the three models, 
where all the neutral CP-even scalars have VEVs.

$O_{\a}$, the mixing matrix for neutral scalars, which is CP-even in nature, is specified as,

\be
 O_\a=
 \begin{pmatrix}
 c_{\a_1} c_{\a_2} & s_{\a_1}c_{\a_2} & s_{\a_2}\cr
 -s_{\a_1}c_{\a_3}-c_{\a_1}s_{\a_2}s_{\a_3} &
 c_{\a_1}c_{\a_3}-s_{\a_1}s_{\a_2}s_{\a_3} & 
 c_{\a_2}s_{\a_3} \cr
 -c_{\a_1}s_{\a_2}c_{\a_3}+s_{\a_1}s_{\a_3} & 
 -c_{\a_1}s_{\a_3}-s_{\a_1}s_{\a_2}c_{\a_3} & c_{\a_2}c_{\a_3}
 \label{eq:Oalpha}
 \end{pmatrix}\,.
\ee

$O_{\b}$, the mixing matrix for charged, and for neutral scalars, which is CP-odd in nature, is specified as,

\be
 O_\b=
 \begin{pmatrix}
  c_{\b} & s_{\b}\cr
  -s_{\b} & c_{\b}
 \label{eq:Obeta} 
 \end{pmatrix}\,.
\ee

In 3HDM, the mixing matrices for CP-odd and charged scalar sector are 
$O_{\b\gamma_1}$ and $O_{\b\gamma_2}$ respectively, with

\be
 O_{\b\gamma_i}=
 \begin{pmatrix}
  c_{\b_1}c_{\b_2} & s_{\b_1}c_{\b_2} & s_{\b_2}\cr
  -s_{\b_1}c_{\gamma_i}-c_{\b_1}s_{\b_2}s_{\gamma_i}
  & c_{\b_1}c_{\gamma_i}-s_{\b_1}s_{\b_2}s_{\gamma_i} 
  & c_{\b_2}s_{\gamma_i}\cr
  -c_{\b_1}s_{\b_2}c_{\gamma_i}+s_{\b_1}s_{\gamma_i}
  & -s_{\b_1}s_{\b_2}c_{\gamma_i}-c_{\b_1}s_{\gamma_i} & c_{\b_2}c_{\gamma_i}
 \end{pmatrix},
 \quad (i=1,2). 
\label{eq:Ogamma} 
\ee

For special cases, where, 

\begin{itemize}
 \item 

either $(i)$ $u_1=0$ in $Rx2SM$, 
or $(ii)$ $v_2=0$ in $N2HDM$, or $(iii)$ $v_2=0$ in $3HDM$, the 
mixing matrix $O_{\a}$ for neutral scalar sector, CP-even in nature as mentioned
in the Eq. (\ref{eq:Oalpha}) is modified to,

\be
 O_\a=
 \begin{pmatrix}
 c_{\a} & 0 & s_{\a}\cr
 0      & 1 & 0      \cr
 -s_{\a}& 0 & c_{\a}
 \end{pmatrix}\,.
\label{eq:Oalpha1}
\ee

\item

either $(i)$ $u_2=0$ in $Rx2SM$, 
or $(ii)$ $u=0$ in $N2HDM$, or $(iii)$ $v_3=0$ in $3HDM$, the 
mixing matrix $O_{\a}$ for neutral scalar sector, CP-even in nature as given
in the Eq. (\ref{eq:Oalpha}) is modified to,

\be
 O_\a=
 \begin{pmatrix}
 c_{\a} & s_{\a} & 0 \cr
 -s_{\a}& c_{\a} & 0 \cr
 0      & 0       & 1
 \end{pmatrix}\,.
\label{eq:Oalpha2}
\ee

\item

either $(i)$ $v_1=0$ in $N2HDM$, or $(ii)$ $v_1=0$ in $3HDM$, the 
mixing matrix $O_{\a}$ for neutral scalar sector, CP-even in nature as given
in the Eq. (\ref{eq:Oalpha}) is modified to,
\be
 O_\a=
 \begin{pmatrix}
 1 & 0 & 0 \cr
 0 & c_{\a} & s_{\a} \cr
 0 & -s_{\a} & c_{\a}
 \end{pmatrix}\,.
\label{eq:Oalpha3}
\ee

\end{itemize}

\section{Oblique parameter calculation in detail}
\label{B}

To calculate the oblique parameters, we require some matix multiplications, 
such as, $Im[({\cal{V}}^{\dag}{\cal{V}})]$, ${\cal{U}}^{\dag}{\cal{V}}$. Here, using Eqns. 
[\ref{eq:UVR1},\ref{eq:UVR2},\ref{eq:UVR3}], we enlist such matrix 
components, useful to derive the results given in the Section 
(\ref{Results}).

\begin{itemize}
 \item Rx2SM
  \be
  Im[{\cal{V}}^{\dag}{\cal{V}}]=
  \begin{pmatrix}
   0 & -O_{{\a}_{11}} & -O_{{\a}_{21}} & -O_{{\a}_{31}} \cr
   O^2_{{\a}_{11}} & 0 & 0 & 0 \cr
   O_{{\a}_{11}} O_{{\a}_{21}} & 0 & 0 & 0 \cr
   O_{{\a}_{11}} O_{{\a}_{31}} & 0 & 0 & 0
  \end{pmatrix}\,,
  \ee
  \be
  {\cal{U}}^{\dag}{\cal{V}}={\cal{V}}=
  \begin{pmatrix}
   i & O_{{\a}_{11}} & O_{{\a}_{21}} & O_{{\a}_{31}}
  \end{pmatrix}\,,\hspace{2cm}
  \ee
  \be
  ({\cal{V}}^{\dag}{\cal{V}})_{11}=1\,,\quad ({\cal{V}}^{\dag}{\cal{V}})_{jj}=O^2_{{\a}_{k1}}\,,
  \,\,{\rm with}\,\, k=(j-1)\,.
  \ee
 \item N2HDM
  \be
  Im[{\cal{V}}^{\dag}{\cal{V}}]= \sum_{l=1}^2
  \begin{pmatrix}
   0 & -O_{{\b}_{1l}}O_{{\a}_{1l}} & 0 & -O_{{\b}_{1l}}O_{{\a}_{2l}} &
   -O_{{\b}_{1l}}O_{{\a}_{3l}}\cr
   O_{{\b}_{1l}}O_{{\a}_{1l}} & 0 & O_{{\b}_{2l}}O_{{\a}_{1l}} &0&0 \cr
   0 & -O_{{\b}_{2l}}O_{{\a}_{1l}} & 0 & -O_{{\b}_{2l}}O_{{\a}_{2l}} 
   & -O_{{\b}_{2l}}O_{{\a}_{3l}}\cr
   O_{{\b}_{1l}}O_{{\a}_{2l}} & 0 & O_{{\b}_{2l}}O_{{\a}_{2l}} &0&0\cr
   O_{{\b}_{1l}}O_{{\a}_{3l}} & 0 & O_{{\b}_{2l}}O_{{\a}_{3l}} &0&0
  \end{pmatrix}\,\,,
  \ee
  \be
  {\cal{U}}^{\dag}{\cal{V}}= \sum_{l=1}^2
  \begin{pmatrix}
   i & O_{{\b}_{1l}}O_{{\a}_{1l}} & 0 & O_{{\b}_{1l}}O_{{\a}_{2l}} &
   O_{{\b}_{1l}}O_{{\a}_{3l}}\cr
   0 & O_{{\b}_{2l}}O_{{\a}_{1l}} & i & O_{{\b}_{2l}}O_{{\a}_{2l}} &
   O_{{\b}_{2l}}O_{{\a}_{3l}}
  \end{pmatrix}\,\,,
  \ee
  \bea
  ({\cal{V}}^{\dag}{\cal{V}})_{11}=1\,,\,\, 
  ({\cal{V}}^{\dag}{\cal{V}})_{22}=\sum_{l=1}^2(O_{{\a}_{1l}})^2\,,\,\,
  ({\cal{V}}^{\dag}{\cal{V}})_{33}=1\,, \n\\
  ({\cal{V}}^{\dag}{\cal{V}})_{44}=\sum_{l=1}^2(O_{{\a}_{2l}})^2\,,\,\,
  ({\cal{V}}^{\dag}{\cal{V}})_{55}=\sum_{l=1}^2(O_{{\a}_{3l}})^2\,.
  \eea
 \item 3HDM
  \be
  Im[{\cal{V}}^{\dag}{\cal{V}}]=\sum_{j=1}^3
  \begin{pmatrix}
   0 & -O_{{\a}_{1j}}O_{{\b\gamma_1}_{1j}} & 0 & 
   -O_{{\a}_{2j}}O_{{\b\gamma_1}_{1j}} & 0 & 
   -O_{{\a}_{3j}}O_{{\b\gamma_1}_{1j}} \cr
   O_{{\a}_{1j}}O_{{\b\gamma_1}_{1j}} & 0 & 
   O_{{\a}_{1j}}O_{{\b\gamma_1}_{2j}} & 0 &
   O_{{\a}_{1j}}O_{{\b\gamma_1}_{3j}} & 0 \cr
   0 & -O_{{\a}_{1j}}O_{{\b\gamma_1}_{2j}} & 0 &
   -O_{{\a}_{2j}}O_{{\b\gamma_1}_{2j}} & 0 &
   -O_{{\a}_{3j}}O_{{\b\gamma_1}_{2j}}\cr
   O_{{\a}_{2j}}O_{{\b\gamma_1}_{1j}} & 0 & 
   O_{{\a}_{2j}}O_{{\b\gamma_1}_{2j}} & 0 &
   O_{{\a}_{2j}}O_{{\b\gamma_1}_{3j}} & 0\cr
   0 & -O_{{\a}_{1j}}O_{{\b\gamma_1}_{3j}} & 0 &
   -O_{{\a}_{2j}}O_{{\b\gamma_1}_{3j}} & 0 &
   -O_{{\a}_{3j}}O_{{\b\gamma_1}_{3j}}\cr
   O_{{\a}_{3j}}O_{{\b\gamma_1}_{1j}} & 0 & 
   O_{{\a}_{3j}}O_{{\b\gamma_1}_{2j}} & 0 &
   O_{{\a}_{3j}}O_{{\b\gamma_1}_{3j}} & 0
  \end{pmatrix}\,,
  \ee
  \bea
  {\cal{U}}^{\dag}{\cal{V}}=&&
  \begin{pmatrix}
   0 & O_{{\b\gamma_2}_{j1}}O_{{\a}_{1j}} & 0 &
   O_{{\b\gamma_2}_{j1}}O_{{\a}_{2j}} & 0 & 
   O_{{\b\gamma_2}_{j1}}O_{{\a}_{3j}}\cr
   0 & O_{{\b\gamma_2}_{j2}}O_{{\a}_{1j}} & 0 &
   O_{{\b\gamma_2}_{j2}}O_{{\a}_{2j}} & 0 & 
   O_{{\b\gamma_2}_{j2}}O_{{\a}_{3j}}\cr
   0 & O_{{\b\gamma_2}_{j3}}O_{{\a}_{1j}} & 0 &
   O_{{\b\gamma_2}_{j3}}O_{{\a}_{2j}} & 
   0 & O_{{\b\gamma_2}_{j3}}O_{{\a}_{3j}}
 \end{pmatrix}\n\\ +\quad i &&
  \begin{pmatrix}
   O_{{\b\gamma_1}_{1j}}O_{{\b\gamma_2}_{j1}} & 0 & 
   O_{{\b\gamma_1}_{2j}}O_{{\b\gamma_2}_{j1}} & 0 & 
   O_{{\b\gamma_1}_{3j}}O_{{\b\gamma_2}_{j1}} & 0\cr
   O_{{\b\gamma_1}_{1j}}O_{{\b\gamma_2}_{j2}} & 0 & 
   O_{{\b\gamma_1}_{2j}}O_{{\b\gamma_2}_{j2}} & 0 & 
   O_{{\b\gamma_1}_{3j}}O_{{\b\gamma_2}_{j2}} & 0\cr
   O_{{\b\gamma_1}_{1j}}O_{{\b\gamma_2}_{j3}} & 0 & 
   O_{{\b\gamma_1}_{2j}}O_{{\b\gamma_2}_{j3}} & 0 & 
   O_{{\b\gamma_1}_{3j}}O_{{\b\gamma_2}_{j3}} & 0
 \end{pmatrix}
 \eea
  \bea
  ({\cal{V}}^{\dag}{\cal{V}})_{jj}&=&1\,,\hspace{1.8cm}{\rm with}\,\, j=(1,3,5)\,,\n\\
  ({\cal{V}}^{\dag}{\cal{V}})_{jj}&=&\sum_{k=1}^3\left({O_{\a_{lk}}}\right)^2\,,
  \,\,{\rm with}\,\, j=(2,4,6)\,,\,\,l=j/2.
  \eea

\end{itemize}


 
\end{document}